\documentclass[aps,prl,twocolumn,showpacs,preprintnumbers]
{revtex4-1}

\usepackage{amssymb}
\usepackage{amsfonts}
\usepackage{amsmath}
\usepackage[dvips]{graphicx}
\usepackage{color}
\usepackage{multirow}
\usepackage{titlesec}
\usepackage{array}

\usepackage{txfonts}
\usepackage{hyperref}
\usepackage{ulem}
\usepackage{setspace}
\definecolor{nred}{rgb}{0.9,0.1,0}

\usepackage{txfonts}

\begin{document}
\title{Deterministic distribution of multipartite entanglement and steering in a quantum network by separable states}

\author{Meihong Wang$^{1,3}$}
\author{Yu~Xiang$^{2,3,4}$}
\author{Haijun Kang$^{1,3}$}
\author{Dongmei Han$^{1,3}$}
\author{Yang Liu$^{1,3}$}
\author{Qiongyi~He$^{2,3,4}$}
\email{qiongyihe@pku.edu.cn}
\author{Qihuang Gong$^{2,3,4}$}
\author{Xiaolong Su$^{1,3}$}
\email{suxl@sxu.edu.cn}
\author{Kunchi Peng$^{1,3}$}

\affiliation{$^1$State Key Laboratory of Quantum Optics and Quantum Optics Devices, Institute
of Opto-Electronics, Shanxi University, Taiyuan, 030006, China \\
$^2$State Key Laboratory for Mesoscopic Physics, School of Physics, Frontiers Science Center for Nano-optoelectronics $\&$ Collaborative Innovation Center of Quantum Matter, Peking University, Beijing 100871, China \\
$^3$Collaborative Innovation Center of Extreme Optics, Shanxi University,
Taiyuan, Shanxi 030006, China \\
$^4$Beijing Academy of Quantum Information Sciences, Beijing 100193, China}

\begin{abstract}
As two valuable quantum resources, Einstein-Podolsky-Rosen entanglement and steering play important roles in quantum-enhanced communication protocols. Distributing such quantum resources among multiple remote users in a network is a crucial precondition underlying various quantum tasks. We experimentally demonstrate the deterministic distribution of two- and three-mode Gaussian entanglement and steering by transmitting separable states in a network consisting of a quantum server and multiple users. In our experiment, entangled states are not prepared solely by the quantum server, but are created among independent users during the distribution process. More specifically, the quantum server prepares separable squeezed states and applies classical displacements on them before spreading out, and users simply perform local beam-splitter operations and homodyne measurements after they receive separable states. We show that the distributed Gaussian entanglement and steerability are robust against channel loss. Furthermore, one-way Gaussian steering is achieved among users that is useful for further directional or highly asymmetric quantum information processing.    
\end{abstract}

\maketitle

Quantum entanglement is an important resource for quantum communication and computation~\cite{RevModPhys}. Besides entanglement, Einstein-Podolsky-Rosen (EPR) steering has also been identified as a valuable resource for secure quantum information tasks~\cite{rmp2020,review2015,reid2009,prxresource}. The states exhibiting steering are a strict subset of the entangled states, and a strict superset of the Bell-nonlocal states~\cite{howard2007}. Distinct from both inseparability and Bell nonlocality, the steerability of two directions between the entangled parties could be asymmetric~\cite{ExpC2009,EFSUV2017} even it can only present in one direction~\cite{OneWayNatPhot}, which has been successfully demonstrated in the pioneer works using continuous variable (CV) Gaussian states~\cite{OneWayNatPhot,ANUexp,prlSu,noiseSu,yinprr}, discrete variable (DV) systems~\cite{NC2015,OneWayPryde,OneWayGuo,XiaoY2017}, and a hybrid CV-DV system~\cite{cvdv}. Remarkably, EPR steering has been created recently in massive~\cite{steeringBEC,steeringAtomic} and high-dimensional systems~\cite{high3,high2,zhangxiangdong,wangjianwei,high1}. The concept of steering is important to quantum networks since it provides a way to verify entanglement, without the trustworthy requirement of the equipment at all nodes of the network. This has abundant 
applications to one-sided device-independent (1SDI) quantum key distribution~\cite{1sDIQKD_howard,HowardOptica,CV-QKDexp}, quantum secret sharing (QSS)~\cite{ANUexp,YuQSS}, secure quantum teleportation~\cite{SQT15,SQT16_LiCM}, and subchannel discrimination~\cite{subchannel,subchannel16}. 

\begin{figure*}[tbph]
	\begin{center}
		\includegraphics[width=150mm]{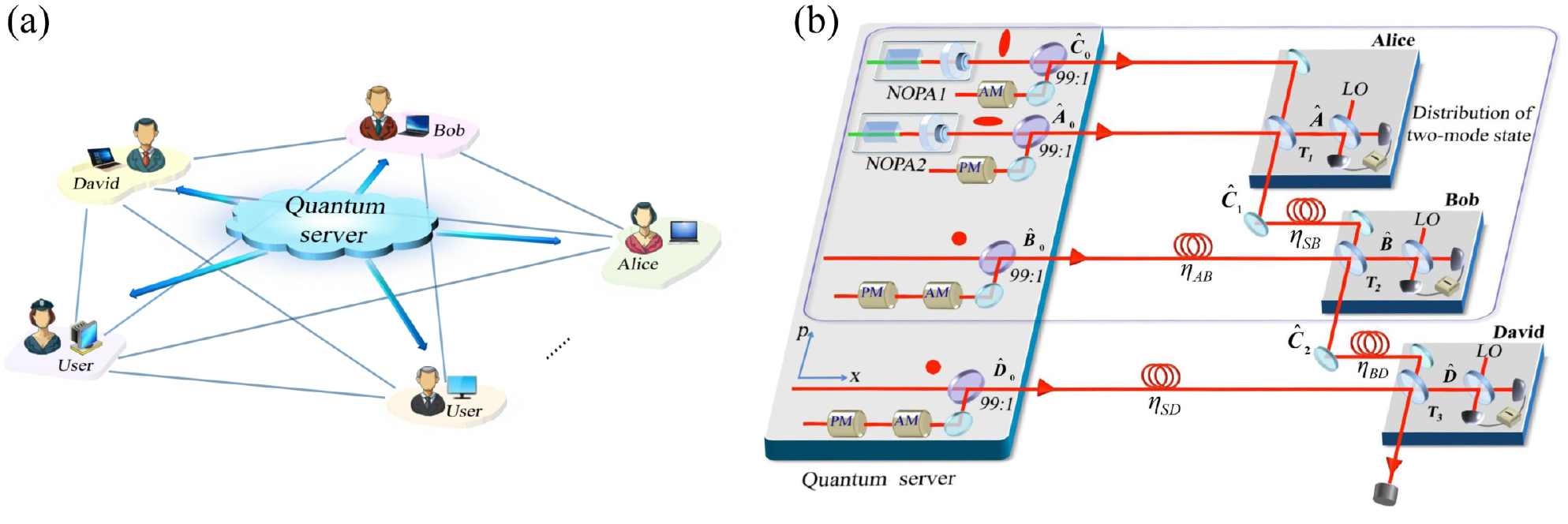}
	\end{center}
	\caption{Schematic of the distribution experiment. $(\text{a})$ Schematic of the quantum network. Quantum server produces quantum states and sends separable states to users. The quantum resource is shared by users after local operations. $(\text{b})$ Schematic of the experimental setup. Two squeezed states with $-3$ dB squeezing ($V_s=0.50$) and $+5.5$ dB antisqueezing ($V_a=3.55$) are produced by two nondegenerate optical parametric amplifiers (NOPA1 and NOPA2). Displacements for all modes are implemented by coupling modulated coherent beams with quantum states on 99:1 beam splitters. The correlated noise is added by amplitude (AM) and phase (PM) modulators, respectively. The distributed states are measured by balanced homodyne detectors for partial reconstruction of the covariance matrix. The lossy channel is simulated by a half-wave plate and a polarization beam splitter.}
	\label{fig1}
\end{figure*}

At the current technology level, it is practical to establish a network consisting of a quantum server, which has the ability to prepare and manipulate quantum states, and two or more users who are merely able to perform local measurements on their states [Fig.~\ref{fig1}(a)]. Consequently, how to distribute entanglement by the quantum server to make it shared among remote users becomes a crucial issue. The conventional method is to directly generate multipartite entangled states by a quantum server locally and then send to remote nodes. Alternatively, there are indirect ways to build entanglement among users, e.g., distributing entanglement by performing joint measurement (entanglement swapping)~\cite{PRLSu2016,otfried network,Miguelnetwork2020}, or by transmitting separable states~\cite{qubitthe,Theory2,Theory3,discord1,discord2,qubitexp,Gaussianexp1,Gaussianexp2}.
In the scheme of distributing entanglement via separable ancilla, instead of preparing entanglement directly by the quantum server, entanglement between two users is created by local operations, classical communication, and transmission of a separable ancillary mode. It has been shown that this indirect method has advantages for distribution of mixed Werner states with depolarizing and dephasing noise~\cite{qubitexp,appnoise}. While significant progress has been reported in recent years~\cite{qubitthe,Theory2,Theory3,discord1,discord2}, as well as first experiments implemented between two qubits~\cite{qubitexp} and between two Gaussian modes~\cite{Gaussianexp1,Gaussianexp2}, the study of this efficient scheme is still in its infancy, and it is fair to say that our understanding of how powerful nonlocality can be provided by this method remains very limited so far. For instance, a generalized scheme was proposed to distribute Gaussian EPR steering by separable states~\cite{YXiang2019}, however, by reanalyzing data from those pioneer experiments~\cite{Gaussianexp1,Gaussianexp2}, we find that none of them were able to demonstrate the shared EPR steering.  As steerability is stronger than inseparability, in general it is harder to distribute steerability than inseparability. Moreover, towards a quantum network, it becomes an even more worthwhile objective to deeply explore the experimental feasibility of distributing multipartite entanglement and steering between more than two users with separable states. In addition, considering the practical channel loss, how to distribute as large as possible steerability at minimal cost is another important problem.

In this Letter, we experimentally demonstrate the deterministic distribution of Gaussian entanglement and steering with separable ancillary states both in two-user and multiuser scenarios. In the experiment, a quantum server prepares independent squeezed states and applies classical displacements on them, which makes initial states fully separable, and then distributes them to users; each user performs a local beam-splitter operation on the received states and transmits one output state of the beam splitter to the next user, where the classical displacements ensure the separability between the transmitted mode and the rest of the states in the network. Instead of providing a particular example to show the entanglement distribution via separable ancilla for two users~\cite{Gaussianexp1,Gaussianexp2}, we rather experimentally implement the distribution of maximal steerability in general by optimizing the displacements according to the initial squeezing level, transmittance of the beam splitter, and transmission efficiencies in the channels. The distributed Gaussian entanglement and steerability are robust against channel loss. Furthermore, moving beyond two parties brings up richer steerability structures including one-way and one-to-multimode steering by mere transmission of separable ancillas, which could be used for providing unprecedented security for a future quantum internet~\cite{kimble08,Wehner}.  

We demonstrate the distribution of multipartite Gaussian entanglement and steering where entangled states are generated deterministically and information is encoded in the position or momentum quadratures of photonic harmonic oscillators~\cite{RevModPhys}. In our experiment, two bright squeezed states are generated by two nondegenerate optical parametric amplifiers (NOPAs). Each of the NOPAs consists of a potassium titanyl phosphate (KTP) crystal and an output coupling mirror. The schematic of the experimental setup is illustrated in Fig.~\ref{fig1}(b), and the details of the experiment can be found in Appendix A. 
The output states are measured in the time domain when the signals of the homodyne detectors are demodulated at a sideband frequency of 3 MHz with a bandwidth of 30 kHz. The demodulated signals are recorded simultaneously by a digital storage oscilloscope at the sampling rate of 500 KS/s.

In the experimental process, a series of correlated displacements (Gaussian noises) need to be optimized and added to realize this indirect distribution. Consequently, much effort is made to make sure the classical noises in same quadratures are canceled at the users' stations, that is, the added Gaussian noises must be synchronized. To do so, all of the displacements added on the amplitude and phase modulators are taken from two independent noise sources, respectively. 
With the increase of the number of users involved in the network, comes the requirement of even more effort to synchronize the added noises on all amplitude and phase modulators. In addition, more relative phases on the beam splitters need to be controlled precisely in the distribution of the three-mode state.

The process for distributing Gaussian entanglement and steering to three users, which is the smallest instance of a true quantum network, contains three steps. In the first step, the quantum server prepares a position (or amplitude quadrature) squeezed state $\hat{C}_{\text{in}}$ and a momentum (or phase quadrature) squeezed state $\hat{A}_{\text{in}}$ generated from two NOPAs, and two coherent (or vacuum) states $\hat{B}_{\text{in}}$ and $\hat{D}_{\text{in}}$ (see Appendix A). 
Then appropriate local classical displacements are applied to all modes according to the following relations:
\begin{eqnarray}
&\hat{x}_{C_{0}} \rightarrow \hat{x}_{C_{\text{in}}}+\mathcal{F}_{C}{x}_{\text{dis}},&\text{ }\hat{p}_{A_{0}}\rightarrow \hat{p}_{A_{\text{in}}}+\mathcal{F}_{A}{p}_{\text{dis}}, \nonumber\\ 
&\hat{x}_{B_{0}} \rightarrow \hat{x}_{B_{\text{in}}}+\mathcal{F}_{B}{x}_{\text{dis}},&\text{ }\hat{p}_{B_{0}}\rightarrow \hat{p}_{B_{\text{in}}}-\mathcal{F}_{B}{p}_{\text{dis}},  \nonumber\\
&\hat{x}_{D_{0}} \rightarrow \hat{x}_{D_{\text{in}}}+\mathcal{F}_{D}{x}_{\text{dis}},&~\hat{p}_{D_{0}}\rightarrow \hat{p}_{D_{\text{in}}}-%
\mathcal{F}_{D}{p}_{\text{dis}},  
\end{eqnarray}%
where $\hat{x}_{j}$ and $\hat{p}_{j}$ represent the position and momentum observables of the state corresponding to the subscript $j$, satisfying the canonical commutation relation $[\hat{x}_{j}, \hat{p}_{j}]=2i$. The classical displacements are determined by ${x}_{\text{dis}}$ and ${p}_{\text{dis}}$ which obey Gaussian distribution with the same variance, and coefficients $\mathcal{F}_{k}$ ($k=A, B, C, D$) corresponding to each mode. The coefficient $\mathcal{F}_{k} (T_{i}, \eta, V_{s, a})$ is a function of transmittance of beam-splitter $T_{i}$, transmission efficiency $\eta$ in the channel, variances of squeezing $V_{s}$ and antisqueezing $V_{a}$ of the input squeezed states. Since $\hat{A}_{\text{in}}$, $\hat{B}_{\text{in}}$, $\hat{C}_{\text{in}}$, $\hat{D}_{\text{in}}$ are prepared independently and the added displacements are local operations and classical communication, the resulting states ${\hat{A}}_{0}$, ${\hat{B}}_{0}$, ${\hat{C}}_{0}$, ${\hat{D}}_{0}$ sent from the quantum server to users are fully separable.

In the second step, optical modes ${\hat{A}}_{0}$ and ${\hat{C}}_{0}$ are transmitted to Alice. We assume that Alice is close to the quantum server, i.e., $\eta_{SA}=1$, while optical modes ${\hat{B}}_{0}$ and ${\hat{D}}_{0}$ are transmitted to Bob and David through lossy channels (the case for $\eta_{SA}\neq1$ is discussed in Appendix E
). In the two-user scenario, only optical mode ${\hat{B}}_{0}$ is transmitted to Bob, while David is not involved.

In the third step, all users perform beam-splitter operations on their received optical modes and measure the obtained states with homodyne detectors. Alice couples modes ${\hat{A}}_{0}$ and ${\hat{C}}_{0}$ on a balanced beam splitter with $T_{1}=1/2$, then keeps one output mode ${\hat{A}}$ and sends the other one ${\hat{C}}_{1}$ to Bob. The displacement operations on initial input modes ensure the separability across ${\hat{C}}_{1}|{\hat{A}\hat{B}}_{0}$ and~$\hat{B}_{0}|{\hat{A}\hat{C}_{1}}$~splittings but entanglement between~$\hat{A}$ and $\hat{B}_{0}\hat{C}_{1}$, which is essential for the present protocol. Bob couples the ancillary mode $\hat{C}_{1}$ and his mode ${\hat{B}}_{0}$ on the beam splitter $T_{2}$. Up to this stage, two-mode entanglement and steering between modes ${\hat{A}}$ and ${\hat{B}}$ (one of the output modes of Bob's beam splitter) are established. Meanwhile, the distributed steerability $\mathcal{G}^{A\rightarrow B}$ can be maximized by optimizing displacement coefficient $\mathcal{F}_B$, which was not uncovered by previous studies. 

In the distribution for three users, the other output mode of Bob's beam splitter ${\hat{C}}_{2}$ is sequentially transmitted to David. A further challenge, apart from the requirement for separability across ${\hat{C}}_{1}|{\hat{A}\hat{B}}_{0}$ splitting, is that we need to carefully design the displacement on mode $\hat{D}_{\text{in}}$ to keep the second ancillary mode $\hat C_2$ separable from all the users' modes $\hat{A}\hat{B}\hat{D}_{0}$. David couples the received mode ${\hat{C}}_{2}$ with his displaced mode $\hat{D}_0$ on the beam splitter $T_{3}$, and hence quantum entanglement and steering among three users, including modes ${\hat{A}}$, ${\hat{B}}$, and ${\hat{D},}$ can be built. Similarly, the Gaussian steerability $\mathcal{G}^{A\rightarrow BD}$ can be maximized by adjusting the displacement coefficient $\mathcal{F}_D$.

The distributed entangled states and the measurements both have Gaussian nature, thus, to detect Gaussian entanglement between subsystems $N$ and $M$ (each subsystem contains $n$ and $m$ modes, respectively) we adopt the positive partial transposition (PPT) criterion~\cite{PPT} which is necessary and sufficient when $n=1$ and $m\geq1$. The separable condition is that all symplectic eigenvalues of the covariance matrix after the partially transposition $\sigma _{NM}^{\top_N}$ are not smaller than $1$ (see Appendix B).

The steerability between two partitions ($N\rightarrow M$) is quantified by the criterion from Ref.~\cite{Kogias2015}, where it was given by 
\begin{equation}
\mathcal{G}^{N\rightarrow M}(\sigma _{NM})=\max \bigg\{0,{-\sum_{j:\bar{\nu}_{j}^{NM\backslash N}<1} }\ln (\bar{\nu}_{j}^{NM\backslash N})\bigg\}.
\end{equation}
Here, $\bar{\nu}_{j}^{NM\backslash N}$ $(j=1,...,m)$ denote the symplectic eigenvalues of the Schur complement $\bar{\sigma}_{NM\backslash N}=\mathcal{M}-\gamma^{\mathsf{T}}\mathcal{N}^{-1}\gamma$ of subsystem $N$, with diagonal blocks $\mathcal{N}$ and $\mathcal{M}$ corresponding to the reduced states of subsystems and the off-diagonal matrices $\gamma$ and $\gamma^{\mathsf{T}}$ encoding the intermodal correlations between subsystems. A nonzero $\mathcal{G}^{N\rightarrow M}>0$ denotes the presence of steering from $N$ to $M$, and a higher value means stronger steerability. The steerability in the opposite direction $\mathcal{G}^{M\rightarrow N}$ can be obtained by swapping the roles of $\mathcal{N}$ and $\mathcal{M}$. The covariance matrices of generated states after each step are detailed in Appendix C. 

\begin{figure}[tbph]
\begin{center}
\includegraphics[width=85mm]{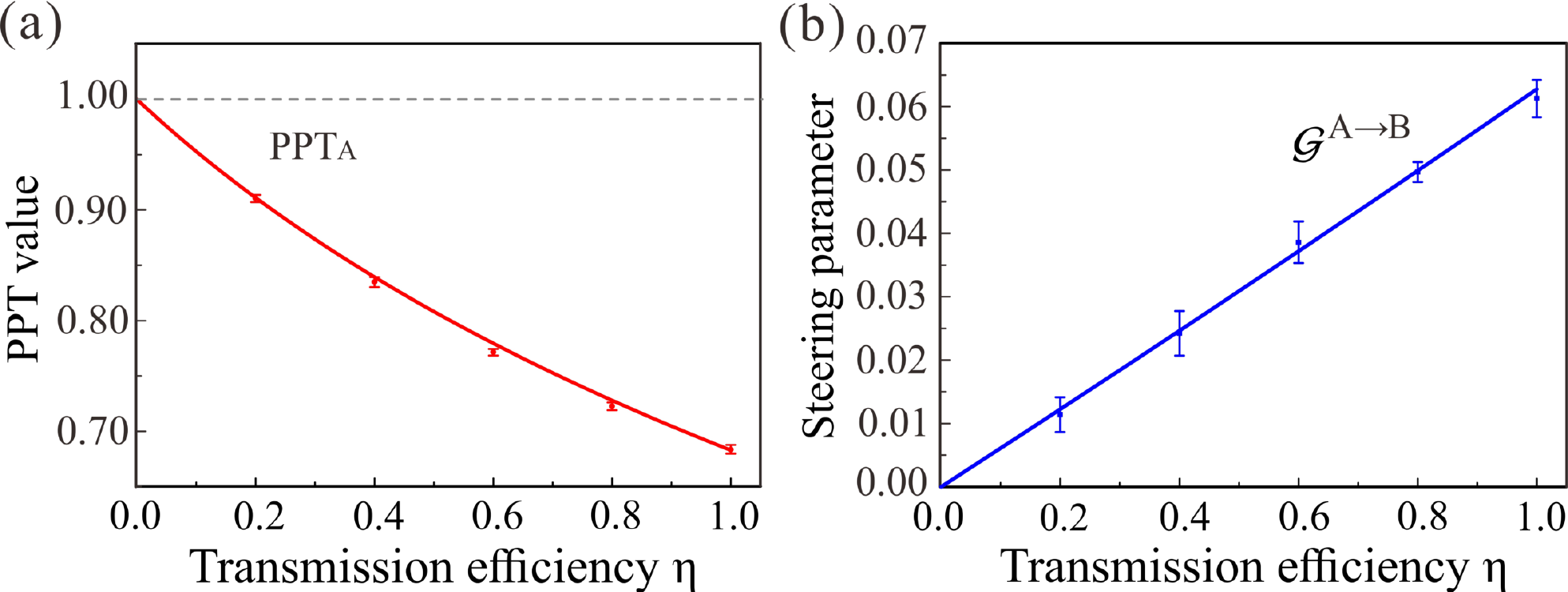}
\end{center}
\caption{Experimental results for two users. $(\text{a})$ The minimum symplectic eigenvalues PPT$_{A}$ with respect to $\hat{A}|\hat{B}$ splitting is always smaller than~$1$. $(\text{b})$ The steerability~$\mathcal{G}^{A\rightarrow B}$ is obtained and robust against loss in channels. Error bars represent one standard deviation and are obtained based on the statistics of measured noise variances.}\label{fig2}
\end{figure}

In this scheme, the crucial idea is that the ancillary modes $({\hat{C}}_{1},{\hat{C}}_{2})$ in the channels are separable from the other modes. The conditions for separability depend on the parameters $\mathcal{F}_{k} (T_{i}, \eta, V_{s, a})$, $x_{\text{dis}}$, and $p_{\text{dis}}$. Without losing generality, we fix the variances of ${x}_{\text{dis}}$ and ${p}_{\text{dis}}$ to $1.50$, $T_{1}=1/2$,~$\mathcal{F}_{A}=\mathcal{F}_{C}=1$, then the condition for separability across ${\hat{C}}_{1}|{\hat{A}\hat{B}}_{0}$ splitting in the two-user scenario only depends on the parameter~$\mathcal{F}_{B}$, and that for the separability across ${\hat{C}_{2}|{\hat{A}\hat{B}}\hat{D}_{0}}$ splitting in the three-user scenario depends on the parameters~$\mathcal{F}_{B}$~and~$\mathcal{F}_{D}$. Additionally, on the basis of satisfying the above separable conditions, we optimize $\mathcal{F}_{k}$ to achieve the highest distributed steerabilities for each desired distribution direction.

To evaluate the performance of the present entanglement and steering distribution network, we investigate the effect of channel loss in our experiment since the transmission distance of the quantum state is limited by inevitable loss in a practical quantum network. In the case of two users, in order to achieve the highest Gaussian steerability $\mathcal{G}^{A\rightarrow B}$, the optimized displacement coefficient on mode~$\hat{B}_{\text{in}}$ is set to $\mathcal{F}_{B}={\sqrt{2\eta_{AB}(1-T_{2})}V_{a}}/\left[{\sqrt{\eta_{SB}T_{2}}(V_{a}+V_{s})}\right]$, where $T_2$ is the transmittance of Bob's beam splitter, and $\eta_{AB}$ and $\eta_{SB}$ are the transmission efficiencies for the channels from Alice to Bob and from quantum server to Bob, respectively. Thus, the maximal distributed steerability $\mathcal{G}^{A\rightarrow B}$ is given by
\begin{equation}
\mathcal{G}^{A\rightarrow B}=\text{ln}\left[\frac{V_{a}+V_{s}}{(1-\eta_{AB}+\eta_{AB}T_{2})(V_{a}+V_{s})+2\eta_{AB}(1-T_{2})V_{s}V_{a}}\right].
\end{equation}

We experimentally fix $T_2=1/2$ and set $\eta_{SB}=\eta_{AB} =\eta$, then the largest distributed steerability $\mathcal{G}^{A\rightarrow B}$ is  
\begin{equation}
\mathcal{G}^{A\rightarrow B}=\text{ln}\left[\frac{2(V_{a}+V_{s})}{(2-\eta
	)(V_{a}+V_{s})+2\eta V_{s}V_{a}}\right] 
\end{equation}
with $\mathcal{F}_{B}\approx1.24$. When all channels are ideal, i.e., $\eta=1$, we measure the covariance matrix~$\sigma_{\text{A}\text{B}_{\text{0}}\text{C}_{\text{1}}}$ and verify the conditions for separability across $\hat{C}_{1}|\hat{A}\hat{B}_{0}$ splitting and~$\hat{B}_{0}|\hat{A}\hat{C}_{1}$~splitting according to their minimum PPT values $1.264>1$ and $1.182>1$, respectively, while~$\hat{A}$~is entangled with group of~$\hat{B}_{0}\hat{C}_{1}$~due to its minimum PPT value $0.701<1$, under the above optimized displacement (see Appendix D). 
Note that when modes~$\hat{C}_{1}$,~$\hat{B}_{0}$~are transmitted in lossy channels, i.e., $\eta<1$, the requirement for the separable conditions will be relaxed. As shown in Fig.~\ref{fig2}, the distributed entanglement between Alice and Bob and one-way Gaussian steerability from Alice to Bob~$\mathcal{G}^{A\rightarrow B}$~always exist when~$\eta>0$~, which means this indirect distribution protocol is robust against channel loss.

After the successful distribution between two users, we extend this protocol to a three-user case. Figure~\ref{fig3} shows that the distributed three-mode entanglement and steerability are also robust against loss in quantum channels. As an example, the transmission efficiencies from quantum server to Bob, quantum server to David, Alice to Bob, and Bob to David are assumed to be the same. To achieve the maximum steerability~$\mathcal{G}^{A\rightarrow BD}$, we optimize the displacements for modes~${\hat{B}}_{{\text{in}}}$~and~${\hat{D}}_{{\text{in}}}$~by~$\mathcal{F}_{B}\approx1.24$~and~$\mathcal{F}_{D}=2\sqrt{\eta}V_{a}/(V_{a}+V_{s})$~with $T_{2}=T_{3}=1/2$. Meanwhile, an additional condition for separability across~$\hat{C}_{2}|\hat{A}\hat{B}\hat{D}_{0}$~splitting needs to be satisfied. Hence, we experimentally reconstruct the covariance matrix~$\sigma _{\text{AB}\text{C}_{\text{2}}\text{D}_{\text{0}}}$~(see Appendix D), 
then verify that the minimum PPT value for splitting across~$\hat{C}_{2}|\hat{A}\hat{B}\hat{D}_{0}$~is~$1.177>1$~when $\eta=1$. Similarly, when modes~$\hat{C}_{2}$,~$\hat{D}_{0}$~are transmitted in lossy channels, the separable condition required by splitting across~$\hat{C}_{2}|\hat{A}\hat{B}\hat{D}_{0}$~is more easily satisfied.
\begin{figure}[htb]
\begin{center}
\includegraphics[width=80mm]{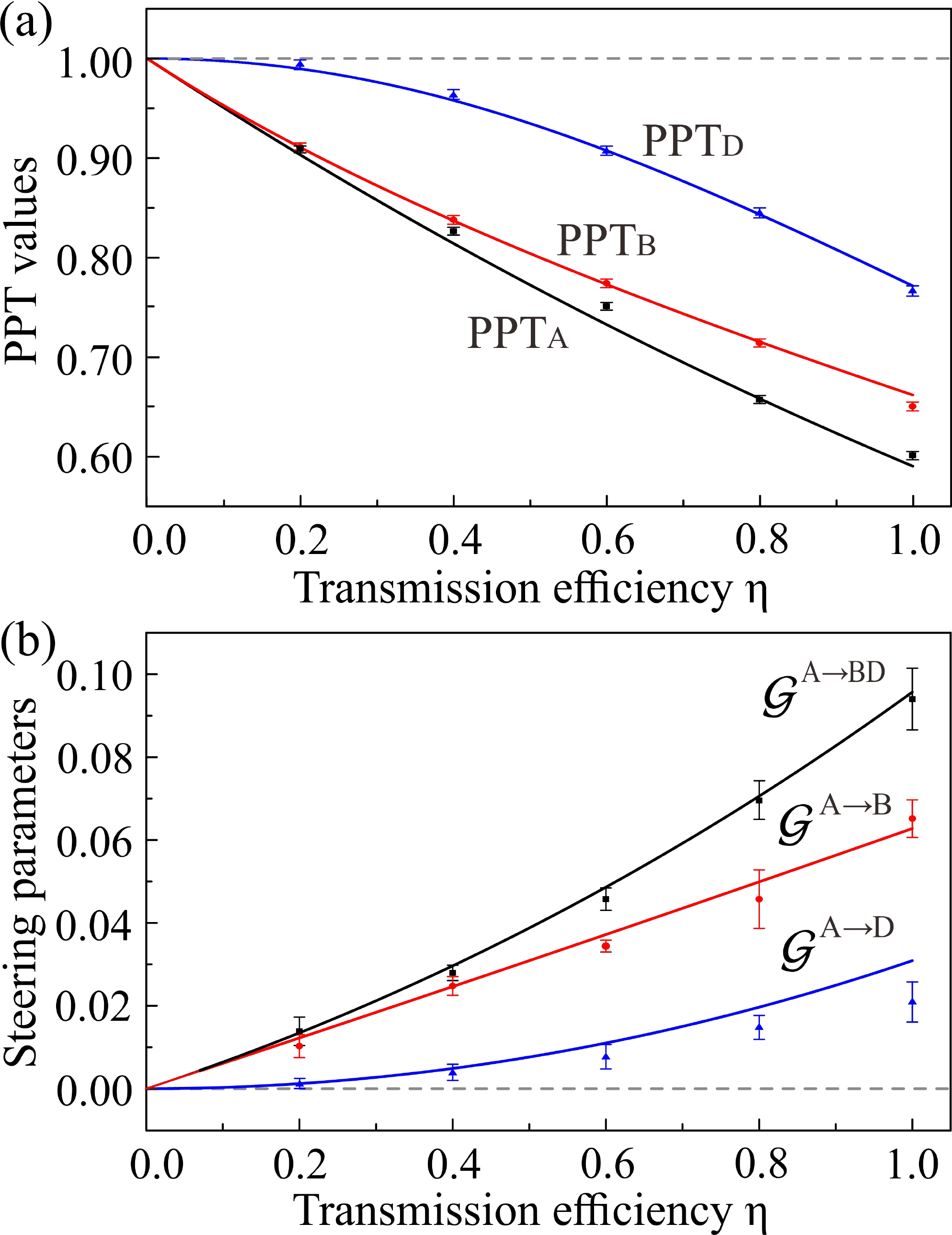}
\end{center}
\caption{Experimental results for three users. 
$(\text{a})$ 
All of the minimum symplectic eigenvalues PPT$_{A}$ (black), PPT$_{B}$ (red), and PPT$_{D}$ (blue) with respect to $\hat{A}|\hat{B}\hat{D}$, $\hat{B}|\hat{A}\hat{D}$, and $\hat{D}|\hat{A}\hat{B}$ splittings are always smaller than~$1$. 
$(\text{b})$ The steerabilities~$\mathcal{G}^{A\rightarrow BD}$, $\mathcal{G}^{A\rightarrow B}$, and $\mathcal{G}^{A\rightarrow D}$ are obtained and robust against channel losses. Error bars represent one standard deviation and are obtained based on the statistics of measured noise variances.}
\label{fig3}
\end{figure}

It is clearly shown in Fig.~\ref{fig3}(a)~that three-mode entanglement is shared among Alice, Bob, and David after the distribution. Different from entanglement, only the one-way steerabilities~$\mathcal{G}^{A\rightarrow BD}>0$,~$\mathcal{G}^{A\rightarrow B}>0$~and~$\mathcal{G}^{A\rightarrow D}>0$~are achieved, and the collective steerability ($\mathcal{G}^{A\rightarrow BD}$) is always higher than the individual steerabilities ($\mathcal{G}^{A\rightarrow B}$~and~$\mathcal{G}^{A\rightarrow D}$), as shown in Fig.~\ref{fig3}(b). We also note that the steering from Bob to David does not exist in any case (i.e.,~$\mathcal{G}^{B\rightarrow D}=0$). This result can be understood as a consequence of the monogamy relation proposed in Ref.~\cite{ReidMI2013} where two independent parties cannot steer a third party simultaneously under Gaussian measurements. Thus,~$\mathcal{G}^{A\rightarrow D}>0$~prohibits the possibility of~$\mathcal{G}^{B\rightarrow D}>0$.
 
Note that the present experimental results show the ability to distribute the Gaussian steerability from Alice to other users (including the individual user and the group of them) by transmitting separable modes. This is because squeezed states are transmitted to Alice firstly, and then separable modes are transmitted from Alice to other users sequentially, i.e., it has a sequential property in such a distribution scheme. It can also be understood in the following way: the final distributed steerability comes from the mixture of two initial squeezed states at Alice's station by a balanced beam splitter, Alice holds half of the information of the whole state, while Bob, David, and the ancillary mode together hold the other half, which makes it much harder for Bob himself, and even together with David, to steer Alice. This means that with the current parameters our experiment presents a highly asymmetric network with directional steerability from Alice to other users. 

After successful distribution, the quantum resources shared among distant users are widely available for real-world applications to networked quantum information tasks. For instance, the hierarchical structure presented in this network, where Alice acts as a superior who can always steer (pilot) any of the subordinate users~($\mathcal{G}^{A\rightarrow B,~D,~BD}>0$), can be applied to implement secure directional quantum key distribution and quantum teleportation from Alice to Bob (David, or their group).

Furthermore, by adjusting the displacements and the transmittances of beam splitters, our protocol can also distribute on-demand quantum resources for specific quantum information tasks. For example, the steerabilities~$\mathcal{G}^{BD\rightarrow A}>0$~and~$\mathcal{G}^{B\rightarrow A}= \mathcal{G}^{D\rightarrow A}=0$~are required for 1SDI QSS, where the dealer Alice sends a secret and players (Bob and David) are able to decode the information only with their collaboration~\cite{YuQSS}. To distribute such a resource via separable ancillas, we need to adjust the displacement coefficients~$\mathcal{F}_{B}=0.92$~and~$\mathcal{F}_{D}=1.70$~with initial~$-10$~dB squeezing and~$+11$~dB antisqueezing, such that the steerability~$\mathcal{G}^{BD\rightarrow A}$~can be distributed for~$0.80< \eta \leq1$~where~$\mathcal{G}^{B\rightarrow A}= \mathcal{G}^{D\rightarrow A}=0$. This means that 1SDI QSS can be implemented in the range of 4.90 km with a fiber loss of 0.2 dB/km~(see Appendix C). 

In summary, we present deterministic distribution of multipartite quantum resources by combining quantum channels and classical communications in a network consisting of a quantum server and multiple users. We demonstrate that it is feasible to distribute not only Gaussian entanglement but also EPR steering among two and three users via separable ancillas. Moreover, the maximum steerability allowed by the present network structure is distributed by optimizing the experimental parameters. The distributed entanglement and steerability are robust against channel losses, which further confirms the significance and practical feasibility of the presented method. This work provides a distinct approach for distributing precious multipartite quantum resources and takes a step forward in studying potential applications of this kind of protocols in a quantum network.

\begin{acknowledgments}
This work was financially supported by National Natural Science Foundation of China (Grants No. 11834010, No. 61675007, No. 11975026, No. 62005149, and No. 12004011), National Key R$\&$D Program of China (Grants No. 2016YFA0301402, No. 2018YFB1107205 and No. 2019YFA0308702). X.S. thanks the program of Youth Sanjin Scholar, and the Fund for Shanxi ``1331 Project" Key Subjects Construction. Q.H. acknowledges the Beijing Natural Science Foundation (Z190005) and the Key R$\&$D Program of Guangdong Province (Grant No. 2018B030329001). 
\end{acknowledgments}

M. W. and Y. X. contributed equally to this work.

\appendix*
\setcounter{equation}{0}
\renewcommand\thefigure{A\arabic{figure}}
\renewcommand\theequation{A\arabic{equation}}
\renewcommand\thetable{A\arabic{table}}
\subsection*{Appendix A:~~~Details of experiment}
\begin{figure*}[tbph]
\begin{center}
\includegraphics[width=150mm]{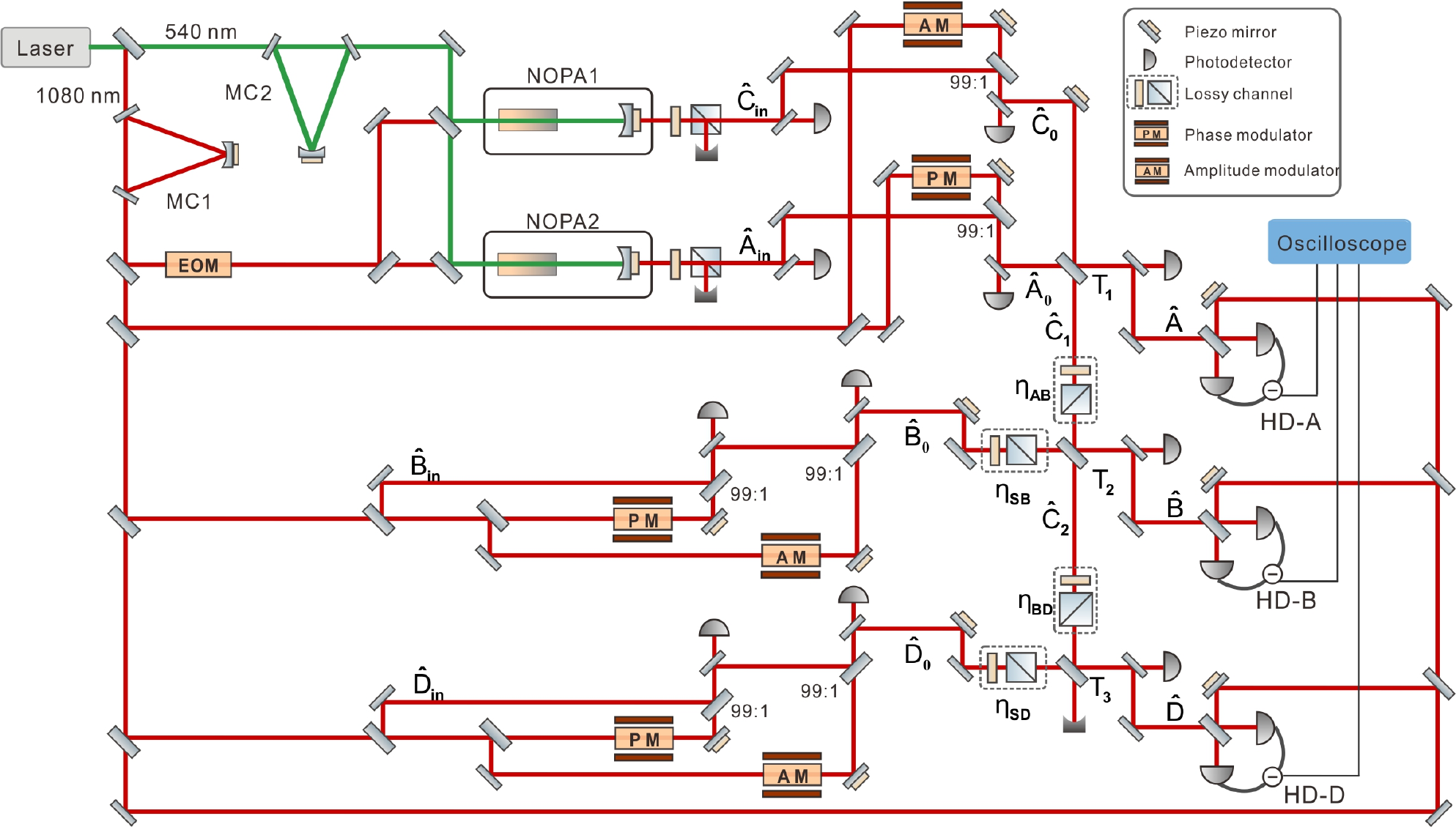}
\end{center}
\setcounter{figure}{0}
\caption{Detailed schematic of experiment setup. MC: mode cleaner. NOPA: non-degenerated optical parametric amplifier. EOM: electro-optical modulator for locking NOPAs (New Focus,~$4004$). PM: phase modulator. AM: amplitude modulator. HD: homodyne detector. HD-A, HD-B and HD-D corresponding to Alice's, Bob's and David's homodyne detector.}
\label{figs1}
\end{figure*}
The experimental setup is shown in Fig.~\ref{figs1}. In our experiment, the $\hat{x}$-squeezed and $\hat{p}$-squeezed states are produced by non-degenerate optical parametric amplifiers (NOPA1 and NOPA2) pumped by a common laser source, which is a continuous wave intracavity frequency-doubled and frequency-stabilized Nd:YAP-LBO (Nd-doped YAlO$_{\text{3}}$ perovskite-lithium triborate) laser. Two mode cleaners are inserted between the laser source and the NOPAs to filter noise and higher order spatial modes of the laser beams at $540$ nm and $1080$ nm. The fundamental wave at $1080$ nm wavelength is used for the injected signals of the NOPAs and the local oscillators for the homodyne detectors. The second-harmonic wave at $540$ nm wavelength serves as pump field of the NOPAs, in which a pair of signal and idler modes with orthogonal polarizations at $1080$ nm are generated through an intracavity frequency-down-conversion process.

The NOPA with a semimonolithic structure, which is similar to that in our previous experiment \cite{NCsu,YZhangPRA,OL8}, is used to generate squeezed states. Each of NOPAs consists of an $\alpha $-cut type-II KTiOPO$_4$ (KTP) crystal ($3\times3\times10~mm^3$) and a concave mirror with curvature~$50$~mm, which is mounted on a piezo-electric transducer (PZT) for locking actively length of NOPAs. The front face of KTP crystal is coated to be used for the input coupler and the concave mirror serves as the output coupler of squeezed states. The transmittances of the front face of KTP crystal at $540$ nm and $1080$ nm are $40\%$ and $0.04\%$, respectively. The end-face of KTP is antireflection coated for both $1080$ nm and $540$ nm. The transmittances of output coupler at $540$ nm and $1080$ nm are $0.5\%$ and $12.5\%$, respectively. The cavity length of each NOPA is~$53.8$~mm. The cavity is locked by using Pound-Drever-Hall method with a phase modulation (New Focus,~$4004$) of~$53$~MHz on injected signal~\cite{PDH}.

When a NOPA is operating at amplification status (the relative phase between injected signal and pump beam is locked to zero), the coupled modes at~$+45^{\circ}$~ and~$-45^{\circ}$~polarization directions are the phase quadrature (momentum) squeezed state and the amplitude quadrature (position) squeezed state, respectively. Conversely, when a NOPA is operating at deamplification status (the relative phase between injected signal and pump beam is locked to~${(2n+1)\pi}$), the coupled modes at $+45^{\circ}$ and $-45^{\circ}$ polarization directions are the amplitude quadrature (position) and the  phase quadrature (momentum) squeezed states, respectively~\cite{NCsu,YZhangPRA,OL8}. 
In the experiment, we choose the bright squeezed state of each NOPA ($+45^{\circ}$~polarization directions) as the signal mode, because the relative phase between squeezed state and assistant coherent beam modulating Gaussian noise needs to be controlled, so NOPA1 and NOPA2 are operating at deamplification and amplification conditions, respectively. 

In the protocol, the displacements for all modes are implemented by the electro-optical modulators. The displacements on the amplitude quadrature are realized by amplitude modulator (AM), and the orthogonality displacements on the phase quadrature are added by phase modulator (PM). The coherent beams carrying the Gaussian noise are mixed with corresponding input mode on~99:1~beam-splitters, and the relative phase differences between the input modes of~99:1~beam-splitters are locked to zero. The displaced modes are coupled on the beam-splitters $T_{1}, T_{2}, T_{3}$ located at the user's stations, and the relative phases of them are also locked to zero. It is important to lock the relative phases of optical modes precisely in the distribution protocol, especially in the case of real networks applications with optical fibers. In order to lock the relative phase between two input modes of each beam-splitter, the transmitted mode of high reflection mirror after the beam-splitter is detected and fed back to a piezo mirror by a microcontroller unit~\cite{MCU}. In our experiment, the phase fluctuation on each beam-splitter is controlled to be around $1^{\circ}$ by adjusting parameters of the phase locking system. The interference efficiencies between two input beams coupled on each beam-splitter are about $99\%$. 

In the distribution of two-mode state, only two displaced squeezed states and a displaced coherent state are transmitted to Alice and Bob, respectively. Alice couples two displaced squeezed states on a beam-splitter $T_{1}$, then keeps one output mode $\hat{A}$ at her station and transmits the other output mode $\hat{C}_{1}$ to Bob. Bob couples modes $\hat{C}_{1}$ and $\hat{B}_{0}$ on a beam-splitter $T_{2}$. One of output mode $\hat{B}$ is obtained and the other output mode $\hat{C}_{2}$ is abandoned. In this way, Gaussian entanglement and steerability between modes~$\hat{A}$~and~$\hat{B}$~are obtained. In the distribution of three-mode state, another displaced coherent state $\hat{D}_{0}$ is also transmitted to David. David couples modes $\hat{C}_{2}$ and $\hat{D}_{0}$ on a beam-splitter $T_{3}$. One of output mode $\hat{D}$
is achieved, and the other output mode $\hat{C}_{3}$ is abandoned. In this way, Gaussian entanglement and steerability among modes $\hat{A}$, $\hat{B}$ and $\hat{D}$ are established. 

For satisfying the separable condition, the variances of ${x}_{\text{dis}}$ and ${p}_{\text{dis}}$ are fixed to~$1.50$, which can be determined according to Eq.~\ref{eqs8a} in the experiment. When the modes $\hat{A}_{\text{in}}$ and~$\hat{C}_{\text{in}}$ are replaced by coherent states ($V_a=V_s=1$ in Eq.~\ref{eqs8a} in this case), we measure variances of amplitude and phase quadratures of mode $\hat{A}$. When measured variances are $2.43$~dB higher than normalized shot noise limit (corresponding to variance of vacuum state), the variances of ${x}_{\text{dis}}$ and ${p}_{\text{dis}}$ will be~$1.50$.

The properties of quantum states are measured by partially reconstructed covariance matrix with the balanced homodyne detectors. The interference efficiencies between signal and local oscillator fields in detection system are~$99\%$~and the quantum efficiencies of photodiodes are $99.6\%$. In our experiment, the output states are measured in the time domain when the signals of homodyne detectors are demodulated at sideband frequency of~$3$ MHz with bandwidth of~$30$~kHz. The demodulated signals are recorded simultaneously by a digital storage oscilloscope at the sampling rate of $500$ KS/s. The squeezing and anti-squeezing levels of the two input states (modes $\hat{A}_{\text{in}}$ and $\hat{C}_{\text{in}}$) are measured by homodyne detectors in the time domain at the quantum server's station, when the displacements are not added.

\subsection*{Appendix B:~~~The criteria of Gaussian entanglement}

The properties of a ($n + m$)-mode Gaussian state of a bipartition system can be determined by its covariance matrix 
\begin{equation}
\sigma _{NM}=\left( 
\begin{array}{cc}
\mathcal{N} & \gamma \\ 
\gamma^{\top } & \mathcal{M}
\end{array}
\right) 
\end{equation}
with matrix element\textit{\ }$\sigma _{ij}=\langle \hat{\xi}_{i}\hat{\xi}%
_{j}+\hat{\xi}_{j}\hat{\xi}_{i}\rangle /2-\langle \hat{\xi}_{i}\rangle
\langle \hat{\xi}_{j}\rangle$\textit{, }where \textit{\ }$\hat{\xi}\equiv (\hat{x}_{1},\hat{p}_{1},...,\hat{x}_{n},\hat{p}_{n},\hat{x}_{n+1},\hat{p}_{n+1}...,\hat{x}_{n+m},\hat{p}_{n+m})^{\intercal}$ \textit{\ }is the vector of the amplitude and phase quadratures of optical modes. The submatrices $\mathcal{N}$ and $\mathcal{M}$ are corresponding to the reduced states of subsystems $N$ and $M$, respectively. 

In the main text, the PPT value is used to quantify the entanglement. Based on above covariance matrix, the symplectic eigenvalue can be calculated. 
If the covariance matrix after partial transposition fulfills the inequality \cite{PPT}
\begin{equation}
\ \sigma _{NM}^{\top_N}+i \Omega \geq{0}
\label{eq5} 
\end{equation}
the state is separable with respect to $N-M$ splitting, where~$\sigma _{NM}^{\top_N}=T_N\sigma _{NM}T_N^{\top}$ is the partially transposed matrix for subsystem~$N$, $T_N$~is a unit diagonal matrix except for the element~$T_{2n,~2n}=-1$, and~$\Omega$~is the symplectic matrix described as
\begin{equation}
\Omega=\oplus_{k=1}^{m+n} \left[
\begin{array}{cc}
 0& 1 \\ 
-1 & 0
\end{array}
\right]  
\end{equation}
This criterion is equivalent to finding the symplectic eigenvalues of the covariance matrix after partial transposition in Ref. \cite{Gaussianexp1}.

When one subsystem holds only one mode~$n=1$~and the other subsystem holds the rest of modes, the state is inseparable if the minimum symplectic eigenvalue of $\sigma _{NM}^{\top_N}$~is smaller than~$1$. For the case of two modes, i.e.,~$m=n=1$, the minimum PPT value (minimum symplectic eigenvalue after partially transpose) is expressed as
\begin{equation}
\mu=\frac{1}{\sqrt{2}}\sqrt{\mathcal{C}-\sqrt{\mathcal{C}^2-4~\text{det} \sigma_{NM}}}
\end{equation}
where~$\mathcal{C}=\text{det} \mathcal{N}+\text{det} \mathcal{M}-2~\text{det} \gamma$.  

\subsection*{Appendix C:~~~The distribution in lossy channels}

\textbf{\newline{The distribution of two-mode state.}}
In the distribution protocol with separable states, it is crucial to make sure the separable condition is maintained in the whole process. In the distribution of two-mode state, the displacement operations applied on initial input modes in the quantum server ensure the separability across ${\hat{C}}_{1}|{\hat{A}\hat{B}}_{0}$ splitting and the inseparability between $\hat{A}$ and ${\hat{B}_{0}\hat{C}_{1}}$. 
After mixing modes~$\hat{A}_{0}$~and~$\hat{C}_{0}$ by beam-splitter $T_{1}$ at Alice's station, the modes~$\hat{A}$~and~$\hat{C}_{1}$~are obtained. Mode ~$\hat{B}_{0}$ is transmitted from quantum server to Bob. After the transmission of optical mode $\hat{o}$ over a lossy channel, the
output optical mode is given by $\hat{o}_{L}=\sqrt{\eta }\hat{o}+\sqrt{%
1-\eta } \hat{o}_{\text{vac}}$,
where $\eta $ and $\hat{o}_{\text{vac}}$ represent the transmission efficiency of lossy channel and optical vacuum mode induced by loss into the quantum channel, respectively. The covariance matrix including modes~$\hat{A}$,~$\hat{B}_{0}$, and~$\hat{C}_{1}$ is given by 
\begin{equation}
\ \ \sigma =\left[ 
\begin{array}{ccc}
\sigma _{A} & \sigma _{AB_{0}} & \sigma _{AC_{1}} \\ 
\sigma _{AB_0}^{\intercal} & \sigma _{B_{0}} & \sigma _{B_{0}C_{1}} \\ 
\sigma _{AC_{1}}^{\intercal} & \sigma _{B_{0}C_{1}}^{\intercal} & \sigma _{C_{1}}
\end{array}\right] 
\end{equation}
where 
\begin{subequations}
\begin{eqnarray}
\ \ \sigma _{A} &=&\left[ 
\begin{array}{cc}
\bigtriangleup ^{2}\hat{x}_{A} & 0 \\ 
0 & \bigtriangleup ^{2}\hat{p}_{A}%
\end{array}%
\right]   \\
\ \sigma _{B_{0}} &=&\left[ 
\begin{array}{cc}
\bigtriangleup ^{2}\hat{x}_{B_{0}} & 0 \\ 
0 & \bigtriangleup ^{2}\hat{p}_{B_{0}}%
\end{array}%
\right]  \\
\ \sigma _{C_{1}} &=&\left[ 
\begin{array}{cc}
\bigtriangleup ^{2}\hat{x}_{C_{1}} & 0 \\ 
0 & \bigtriangleup ^{2}\hat{p}_{C_{1}}%
\end{array}%
\right]  
\end{eqnarray}
\begin{eqnarray}
\sigma _{AB_{0}} &=&\left[ 
\begin{array}{cc}
Cov(\hat{x}_{A},\hat{x}_{B_{\text{0}}}) & Cov(\hat{x}_{A},\hat{p}_{B_{\text{0}}}) \\ 
Cov(\hat{p}_{A},\hat{x}_{B_{\text{0}}}) & Cov(\hat{p}_{A},\hat{p}_{B_{\text{0}}})%
\end{array}\right]  \\
\sigma _{AC_{1}} &=&\left[ 
\begin{array}{cc}
Cov(\hat{x}_{A},\hat{x}_{C_{1}}) & Cov(\hat{x}_{A},\hat{p}_{C_{1}}) \\ 
Cov(\hat{p}_{A},\hat{x}_{C_{1}}) & Cov(\hat{p}_{A},\hat{p}_{C_{1}})%
\end{array}\right]  \\
\sigma _{B_{0}C_{1}} &=&\left[ 
\begin{array}{cc}
Cov(\hat{x}_{B_{0}},\hat{x}_{C_{1}}) & Cov(\hat{x}_{B_{0}},\hat{p}_{C_{1}}) \\ 
Cov(\hat{p}_{B_{0}},\hat{x}_{C_{1}}) & Cov(\hat{p}_{B_{0}},\hat{p}_{C_{1}})%
\end{array}\right]
\end{eqnarray}
\end{subequations}
The elements in the matrices are given by
\begin{subequations}
\begin{eqnarray}
\ \bigtriangleup ^{2}\hat{x}_{A} &=&\bigtriangleup ^{2}\hat{p}_{A}=\frac{%
V_{a}+V_{s}+V_{\text{dis}}}{2}  
 \label{eqs8a}\\
\bigtriangleup ^{2}\hat{x}_{B_{0}} &=&\bigtriangleup ^{2}\hat{p}_{B_{0}}=\eta_{SB}(1+V_{\text{dis}}\mathcal{F}_{B}^2)+1-\eta_{SB} \\
\bigtriangleup ^{2}\hat{x}_{C_{\text{1}}} &=&\frac{\eta_{AB} (V_{a}+V_{s}+V_{\text{dis}})}{2}+1-\eta_{AB} \\ 
Cov(\hat{x}_{A},\hat{x}_{B_{0}}) &=&-Cov(\hat{p}_{A},\hat{p}_{B_{0}})=\frac{\sqrt{2\eta_{SB}}V_{\text{dis}}\mathcal{F}_{B}}{2}  \\
Cov(\hat{x}_{A},\hat{x}_{C_{1}}) &=&-Cov(\hat{p}_{A},\hat{p}_{C_{1}})=\frac{\sqrt{\eta_{AB}}(V_{a}-V_{s}-V_{\text{dis}})}{2}  \notag \\
   \\
Cov(\hat{x}_{B_{0}},\hat{x}_{C_{1}}) &=&Cov(\hat{p}_{B_{0}},\hat{p}_{C_{1}})=-\frac{\sqrt{2\eta_{AB}\eta_{SB}}V_{\text{dis}}\mathcal{F}_{B}}{2}   \\
Cov(\hat{x}_{A},\hat{p}_{B_{0}})&=&Cov(\hat{p}_{A},\hat{x}_{B_{0}})=0 \\
Cov(\hat{x}_{A},\hat{p}_{C_{1}})&=&Cov(\hat{p}_{A},\hat{x}_{C_{1}})=0 \\                                                                                
Cov(\hat{x}_{B_{0}},\hat{p}_{C_{1}})&=&Cov(\hat{p}_{B_{0}},\hat{x}_{C_{1}})=0                                     
\end{eqnarray}
\end{subequations}
where $V_{a}$,~$V_{s}$ are the variances of squeezing and anti-squeezing of initial squeezed states, $V_{\text{dis}}=\langle(\Delta {x}_{\text{dis}})^2\rangle=\langle(\Delta {p}_{\text{dis}})^2\rangle$ is the variance of displacements on modes~$\hat{A}_{\text{in}}$~and~$\hat{C}_{\text{in}}$,~$\mathcal{F}_{B}$~is the corresponding coefficient for mode~$\hat{B}_{\text{in}}$,~and~$\eta_{AB}$~and~$\eta_{SB}$~are the transmission efficiencies from Alice to Bob and that from quantum server to Bob, respectively. 

\begin{figure*}[tbph]
\begin{center}
\includegraphics[width=150mm]{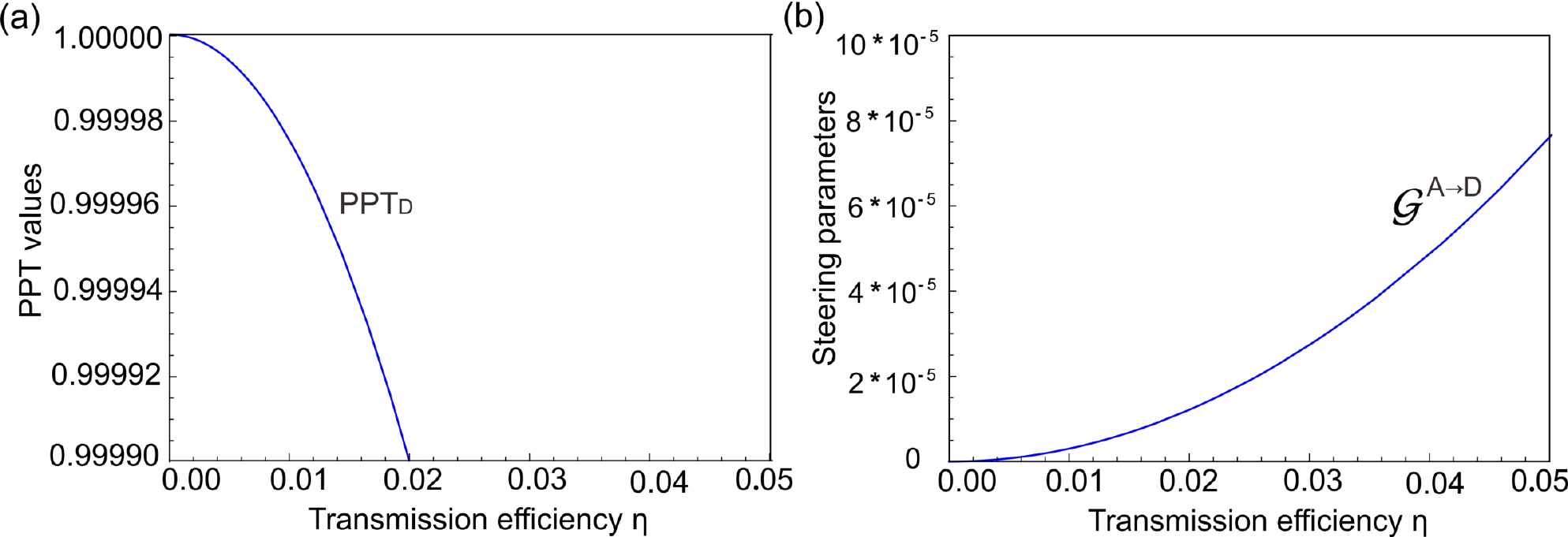}
\end{center}
\caption{Details of Fig. 3 in the main text. 
\text{(a)} PPT value PPT$_{D}$~for the splitting~$\hat{D}|\hat{A}\hat{B}$ and \text{(b)} steering parameter $\mathcal{G}^{A\rightarrow D}$ for the very small value of transmission efficiencies in the distribution of three-mode state, when the displacement coefficients~$\mathcal{F}_{B}$~and~$\mathcal{F}_{D}$~are optimized.} \label{figs2}
\end{figure*}

In the distribution of two-mode state, the separability across~$\hat{C}_{1}|\hat{A}\hat{B}_{0}$~splitting and inseparability between~$\hat{A}$~and the group~($\hat{B}_{0}\hat{C}_{1}$)~are required. By classical displacement operations, the separability across splitting~$\hat{C}_{1}|\hat{A}\hat{B}_{0}$~is satisfied when
\begin{equation}
V_{\text{dis}}>V_{\text{sep}}=\frac{2(1-V_{s})}{2-\eta_{SB}\mathcal{F}_{B}^2(1-V_{s})} 
\label{eqs4}
\end{equation} 
where the minimum requirement for the variance of displacement $V_{\text{sep}}$ is named as the separable boundary. The inseparable condition for~$\hat{A}|\hat{B}_{0}\hat{C}_{1}$~splitting is $V_{\text{dis}}>0$, which is more relax than separable boundary $V_{\text{sep}}$. Therefore, the classical displacements on modes~$\hat{A}_{\text{in}}$~and~$\hat{C}_{\text{in}}$~should be larger than~$V_{\text{sep}}$~in order to satisfy the separable condition. 

After the ancillary mode~$\hat{C}_{1}$~is transmitted to Bob, the modes~$\hat{B}_{0}$~and~$\hat{C}_{1}$~are coupled on Bob's beam-splitter~$T_{2}$. Thus, the covariance matrix of the output modes $\hat{A}$ and $\hat{B}$ in the distribution between two users is expressed by
\begin{equation}
\ \ \sigma =\left[ 
\begin{array}{ccc}
\sigma _{A} & \sigma _{AB} \\ 
\sigma _{AB}^{\intercal} & \sigma _{B}  \\ 
\end{array}%
\right] 
\end{equation}
where 
\begin{subequations}
\begin{eqnarray}
\ \ \sigma _{A} &=&\left[ 
\begin{array}{cc}
\bigtriangleup ^{2}\hat{x}_{A} & 0 \\ 
0 & \bigtriangleup ^{2}\hat{p}_{A}%
\end{array}%
\right]    \\
\ \sigma _{B} &=&\left[ 
\begin{array}{cc}
\bigtriangleup ^{2}\hat{x}_{B} & 0 \\ 
0 & \bigtriangleup ^{2}\hat{p}_{B}%
\end{array}%
\right]   \\
\sigma _{AB} &=&\left[ 
\begin{array}{cc}
Cov(\hat{x}_{A},\hat{x}_{B}) & Cov(\hat{x}_{A},\hat{p}_{B}) \\ 
Cov(\hat{p}_{A},\hat{x}_{B}) & Cov(\hat{p}_{A},\hat{p}_{B})%
\end{array}%
\right] 
\end{eqnarray}
\end{subequations}
The elements in the matrices are given by
\begin{subequations}
\begin{eqnarray}
\ \bigtriangleup ^{2}\hat{x}_{A} &=&\bigtriangleup ^{2}\hat{p}_{A}=\frac{%
V_{a}+V_{s}+V_{\text{dis}}}{2} \\
\bigtriangleup ^{2}\hat{x}_{B} &=&\bigtriangleup ^{2}\hat{p}_{B}  \notag \\
                                               &=&\frac{\eta_{AB}(1-T_{2})(V_{a}+V_{s}+V_{\text{dis}})}{2}+\eta_{SB}T_{2}V_{\text{dis}}\mathcal{F}_{B}^2 \notag \\
&-&\sqrt{2\eta_{SB}\eta_{AB}T_{2}(1-T_{2})}V_{\text{dis}}\mathcal{F}_{B} +1-\eta_{AB}+\eta_{AB} T_{2}   \notag \\
 \\
Cov(\hat{x}_{A},\hat{x}_{B}) &=&-Cov(\hat{p}_{A},\hat{p}_{B}) \notag \\
                                             &=&\frac{\sqrt{\eta_{AB}(1-T_{2})}(V_{a}-V_{s}-V_{\text{dis}})+\sqrt{2\eta_{SB} T_{2}}V_{\text{dis}}\mathcal{F}_{B}}{2} \notag \\
                                             \\
                                             Cov(\hat{x}_{A},\hat{p}_{B})&=&Cov(\hat{p}_{A},\hat{x}_{B})=0
\end{eqnarray}
\end{subequations}

In order to obtain the maximum steerability~$\mathcal{G}^{A\rightarrow B}$, the displacement coefficient~$\mathcal{F}_{B}$~needs to be optimized. Here, the optimal~$\mathcal{F}_{B}={\sqrt{2\eta_{AB}(1-T_{2})}V_{a}}/[{\sqrt{\eta_{SB}T_{2}}(V_{a}+V_{s})}]$, which is the function of transmittance~$T_{2}$, transmission efficiencies~$\eta_{SB}$,~$\eta_{AB}$, and the variances of squeezing and anti-squeezing of squeezed states. By substituting optimized coefficient~$\mathcal{F}_{B}$~into Eq. \ref{eqs4}, the separable boundary across~$\hat{C}_{1}|\hat{A}\hat{B}_{0}$~splitting between two users is expressed by  
\begin{equation}
\ V_{\text{sep}}=\frac{T_{2}(1-V_{s})(V_{a}+V_{s})^2}{T_{2}(V_{a}+V_{s})^2-\eta_{AB}(1-T_{2})(1-V_{s})V_{a}^2} \label{eqs13}
\end{equation}
From this expression, we can see that the separable boundary decreases with the decrease of transmission efficiency $\eta_{AB}$ from Alice to Bob, when the rest of parameters are fixed. When the transmittance of Bob's beam-splitter is chosen to~$T_{2}=1/2$~and transmission efficiency~$\eta_{AB}=\eta_{SB}=\eta=1$, the separable boundary $V_{\text{sep}}=0.808$. In our experiment, we choose~$V_{\text{dis}}=1.50$~which satisfies the separable condition. 

The dependence of maximum steerability $\mathcal{G}%
^{A\rightarrow B}$ on the transmission efficiency~$\eta$~in the
distribution of two-mode state through lossy channels (blue curve in Fig. 2 in the main text) is expressed by
\begin{equation}
\mathcal{G}^{A\rightarrow B}=\text{ln}[\frac{2(V_{a}+V_{s})}{(2-\eta
)(V_{a}+V_{s})+2\eta V_{s}V_{a}}] 
\end{equation}
where the transmittances $T_{1}=T_{2}=1/2$ and optimal displacement $\mathcal{F}_{B}={\sqrt{2}V_{a}}/{(V_{a}+V_{s})}$ are chosen.

\textbf{\newline{The distribution of three-mode state.}} The output state covariance matrix in case of the distribution between three users, in the presence of lossy channels can be expressed as

\begin{equation}
\ \ \sigma =\left[ 
\begin{array}{ccc}
\sigma _{A} & \sigma _{AB} & \sigma _{AD} \\ 
\sigma _{AB}^{\intercal} & \sigma _{B} & \sigma _{BD} \\ 
\sigma _{AD}^{\intercal} & \sigma _{BD}^{\intercal} & \sigma _{D}%
\end{array}%
\right] 
\end{equation}%
where
\begin{subequations}
\begin{eqnarray}
\ \ \sigma _{A} &=&\left[ 
\begin{array}{cc}
\bigtriangleup ^{2}\hat{x}_{A} & 0 \\ 
0 & \bigtriangleup ^{2}\hat{p}_{A} 
\end{array}
\right]  \\
\sigma _{B} &=&\left[ 
\begin{array}{cc}
\bigtriangleup ^{2}\hat{x}_{B} & 0 \\ 
0 & \bigtriangleup ^{2}\hat{p}_{B}%
\end{array}%
\right]  \\
\sigma _{D} &=&\left[ 
\begin{array}{cc}
\bigtriangleup ^{2}\hat{x}_{D} & 0 \\ 
0 & \bigtriangleup ^{2}\hat{p}_{D}%
\end{array}%
\right]  \\
\sigma _{AB} &=&\left[ 
\begin{array}{cc}
Cov(\hat{x}_{A},\hat{x}_{B}) & Cov(\hat{x}_{A},\hat{p}_{B}) \\ 
Cov(\hat{p}_{A},\hat{x}_{B}) & Cov(\hat{p}_{A},\hat{p}_{B})%
\end{array}%
\right]   \\
\sigma _{AD} &=&\left[ 
\begin{array}{cc}
Cov(\hat{x}_{A},\hat{x}_{D}) & Cov(\hat{x}_{A},\hat{p}_{D}) \\ 
Cov(\hat{p}_{A},\hat{x}_{D}) & Cov(\hat{p}_{A},\hat{p}_{D})%
\end{array}%
\right]   
\end{eqnarray}
\begin{eqnarray}
\sigma _{BD} &=&\left[ 
\begin{array}{cc}
Cov(\hat{x}_{B},\hat{x}_{D}) & Cov(\hat{x}_{B},\hat{p}_{D}) \\ 
Cov(\hat{p}_{B},\hat{x}_{D}) & Cov(\hat{p}_{B},\hat{p}_{D})%
\end{array}%
\right]
\end{eqnarray}
\end{subequations}
\begin{table}[bp]
\renewcommand\arraystretch{1.2}
\begin{tabular}{|p{13mm}<{\centering}|p{13mm}<{\centering}|p{13mm}<{\centering}|p{13mm}<{\centering}|p{13mm}<{\centering}|p{13mm}<{\centering}|}
\multicolumn{6}{c}{\textbf{Table A1} \textbf{Optimal displacements in lossy channels.\bigskip}}   \\ 
\hline
$\eta$&$\textbf{1}$&$\textbf{0.8}$&$\textbf{0.6}$&$\textbf{0.4}$&$\textbf{0.2}$ \\ \hline
$\mathcal{F}_{B}$&$1.239$&$1.239$&$1.239$&$1.239$&$1.239$  \\ \hline 
$\mathcal{F}_{D}$&$1.752$&$1.567$&$1.357$&$1.108$&$0.784$   \\ \hline
\end{tabular}
\label{tables2}
\end{table}
The elements in the matrices are given by
\begin{subequations}
\begin{eqnarray}
\ \bigtriangleup ^{2}\hat{x}_{A} &=&\bigtriangleup ^{2}\hat{p}_{A}=\frac{%
V_{a}+V_{s}+V_{\text{dis}}}{2}   \\
\bigtriangleup ^{2}\hat{x}_{B} &=&\bigtriangleup ^{2}\hat{p}_{B}=\frac{\eta(V_{a}+V_{s}+V_{\text{dis}})}{4}+f   
\end{eqnarray}
\begin{eqnarray}
\bigtriangleup ^{2}\hat{x}_{D} &=&\bigtriangleup ^{2}\hat{p}_{D}=\frac{\eta^{2}(V_{a}+V_{s}+V_{\text{dis}})}{8}+g \\
Cov(\hat{x}_{A},\hat{x}_{B}) &=&-Cov(\hat{p}_{A},\hat{p}_{B}) \notag\\
&=&\frac{\sqrt{2\eta }(V_{a}-V_{s}-V_{\text{dis}}+\sqrt{2}V_{\text{dis}}\mathcal{F}_{B})}{4} \notag\\
\\
Cov(\hat{x}_{A},\hat{x}_{D}) &=&-Cov(\hat{p}_{A},
\hat{p}_{D}) \notag\\
&=&\frac{\eta (V_{a}-V_{s}-V_{\text{dis}})}{4}+j  \\
Cov(\hat{x}_{B},\hat{x}_{D})&=&Cov(\hat{p}_{B},
\hat{p}_{D})  \notag\\
&=&\frac{\sqrt{2\eta ^{3}}(V_{a}+V_{s}+V_{\text{dis}})}{8}+k \\
Cov(\hat{x}_{A},\hat{p}_{B})&=&Cov(\hat{p}_{A},\hat{x}_{B})=0 \\
Cov(\hat{x}_{A},\hat{p}_{D})&=&Cov(\hat{p}_{A},\hat{x}_{D})=0  \\
Cov(\hat{x}_{B},\hat{p}_{D})&=&Cov(\hat{p}_{B},\hat{x}_{D})=0 
\end{eqnarray}
\end{subequations}
where
\begin{subequations}
\begin{eqnarray}
f&=&\frac{\eta }{2}(V_{\text{dis}}\mathcal{F}_{B}^{2}-1-\sqrt{2}V_{\text{dis}}\mathcal{F}_{B})+1  \\
g&=&\frac{4+\eta ^{2}(V_{\text{dis}}\mathcal{F}_{B}^{2}+\sqrt{2}V_{\text{dis}}\mathcal{F}_{B}-1)+2\eta V_{\text{dis}}\mathcal{F}_{D}^{2}}{4} \notag \\
  &-&\frac{2\sqrt{\eta ^{3}}V_{\text{dis}}\mathcal{F}_{D}(\sqrt{2}\mathcal{F}_{B}+1)}{4} \\
j&=&\frac{2\sqrt{\eta }V_{\text{dis}}\mathcal{F}_{D}-\sqrt{2}\eta V_{\text{dis}}\mathcal{F}_{B}}{4}  \\
k&=&\frac{-\sqrt{2\eta ^{3}}(V_{\text{dis}}\mathcal{F}_{B}^{2}+1)+\sqrt{2}\eta V_{\text{dis}}\mathcal{F}_{D}(\sqrt{2}\mathcal{F}_{B}-1)}{4} \notag \\
\end{eqnarray}
\end{subequations}
and~$\eta=\eta_{SB}=\eta_{AB}=\eta_{SD}=\eta_{BD}$~is the transmission efficiency of all lossy channels, 
~$\mathcal{F}_{D}$ is the displacement coefficient for mode~$\hat{D}_{\text{in}}$.
\begin{figure}[hbph]
\begin{center}
\includegraphics[width=80mm]{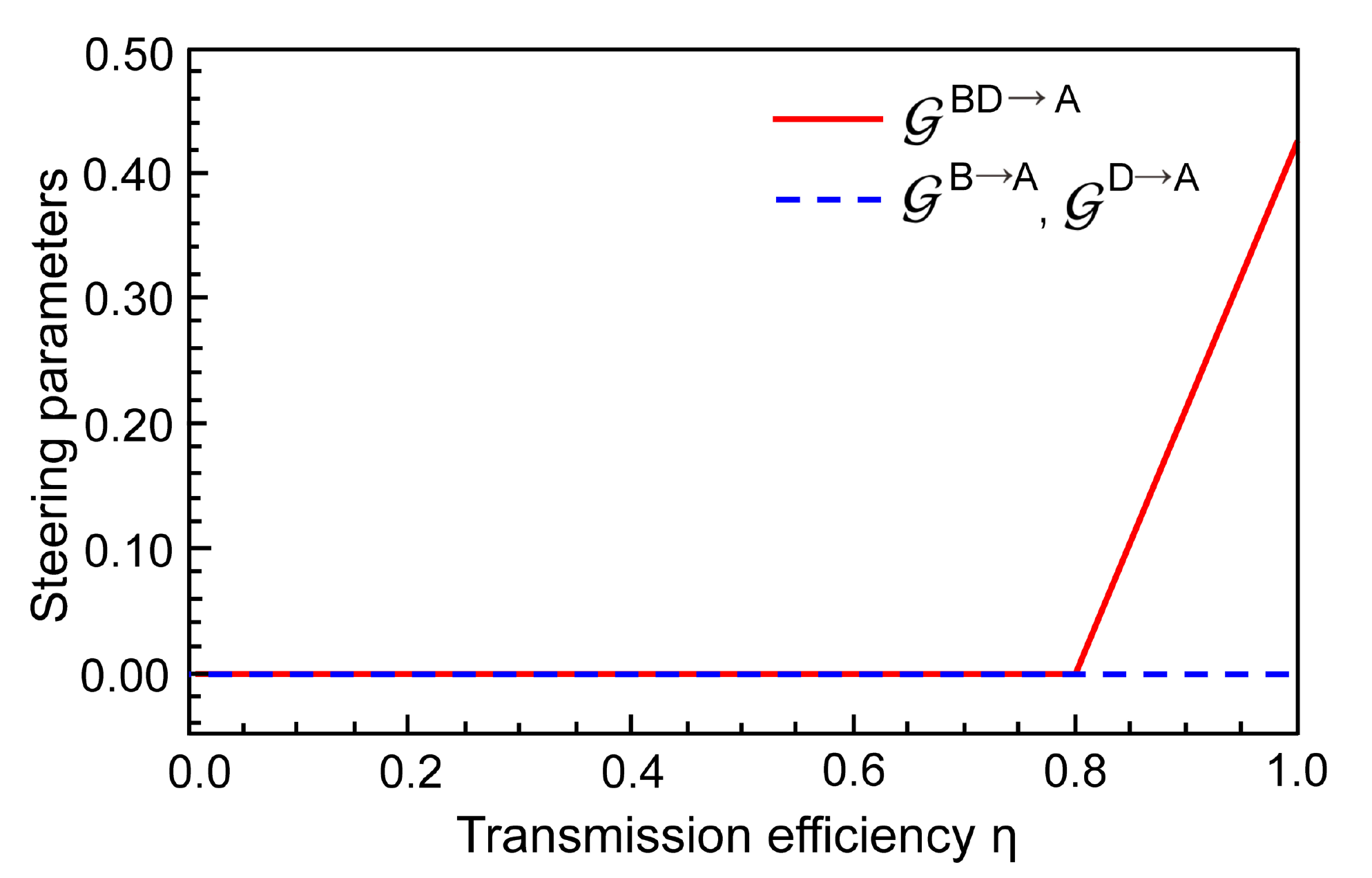}
\end{center}
\caption{The dependences of steerabilities~$\mathcal{G}^{BD\rightarrow A}$,~$\mathcal{G}^{B\rightarrow A}$~and~$\mathcal{G}^{D\rightarrow A}$~on transmission efficiency. The steerability~$\mathcal{G}^{BD\rightarrow A}>0$ when the transmission efficiency is large than $0.80$, and neither Bob nor David has ability to steer Alice.} \label{figs3}
\end{figure}

When three users are involved in the quantum network and~$T_{1},~T_{2}$,~$T_{3}$~are fixed to 1/2, the optimal displacements for modes $\hat{B}_{\text{in}}$ and $\hat{D}_{\text{in}}$ are
optimized as~$\mathcal{F}_{B}={\sqrt{2}V_{a}}/{(V_{a}+V_{s})}$~and~$\mathcal{F}_{D}={2\sqrt{\eta}V_{a}}/{(V_{a}+V_{s})}$ for achieving the maximun
steerability $\mathcal{G}^{A\rightarrow BD}$ in lossy channels. Hence the
steerabilities $\mathcal{G}^{A\rightarrow BD},~\mathcal{G}^{A\rightarrow B},$
and $\mathcal{G}^{A\rightarrow D}$ are expressed by
\begin{subequations}
\begin{eqnarray}
\ \mathcal{G}^{A\rightarrow BD}&=&\text{ln}[\frac{4(V_{a}+V_{s})}{(4-\eta ^{2}-2\eta
)(V_{a}+V_{s})+(4\eta +2\eta ^{2})V_{s}V_{a}}]  \notag\\
\\
\ \mathcal{G}^{A\rightarrow B}&=&\text{ln}[\frac{2(V_{a}+V_{s})}{(2-\eta
)(V_{a}+V_{s})+2\eta V_{s}V_{a}}]   \\
\ \ \mathcal{G}^{A\rightarrow D}&=&\text{ln}[\frac{4(V_{a}+V_{s})}{(4-\eta
^{2})(V_{a}+V_{s})+2\eta ^{2}V_{s}V_{a}}] 
\end{eqnarray}
\end{subequations}
respectively. 

The optimal displacements in the presence of lossy channel with different transmission efficiencies $\eta $ are listed in the Table A1. We measure all the covariance matrices of the distributed modes experimentally, and then calculate the steerability and PPT values to quantify the EPR steering and entanglement. In the distribution of three-mode state, the blue curves of Fig. 3 in the main text are shown in Fig. \ref{figs2} for very small value of transmission efficiencies, which clearly shows that PPT value for splitting~$\hat{D}|\hat{A}\hat{B}$~and steerability~$\mathcal{G}^{A\rightarrow D}$ are robust against loss.

Furthermore, our protocol can be also applied to distribute steerability in the opposite direction that $\mathcal{G}^{BD\rightarrow A}>0$ while $\mathcal{G}^{B\rightarrow A}=\mathcal{G}^{D\rightarrow A}=0$ by appropriately adjusting experimental parameters. The three-mode state with this steering properties is a necessary resource for one-sided device-independent quantum secret sharing (1sDI QSS), a protocol used to send a highly important message to two players (need not assume reliable devices) who must collaborate to obtain the information sent by the dealer. To distribute such resource via separable ancillas, we keep the transmittances of all beam-splitters ~$T_{1}=T_{2}=T_{3}=1/2$~and displacements~$x_{\text{dis}}=p_{\text{dis}}=1.50$ unchanged, and then choose optimal displacement coefficients~$\mathcal{F}_{B}=0.92$~and~$\mathcal{F}_{D}=1.70$ for the initial $-10$ dB squeezing and~$+11$~dB anti-squeezing to distribute as large as possible steerability from the secrete receivers $\hat{B}\hat{D}$ to the dealer $\hat{A}$. Simultaneously, the separable conditions during distribution process are guaranteed, which are evidenced by the minimum PPT values for splittings across~$\hat{C}_{1}|\hat{A}\hat{B}_{0}$~and~$\hat{C}_{1}|\hat{A}\hat{B}\hat{D}_{0}$ being $1.02>1$ and $1.01>1$, respectively.

From Fig. \ref{figs3} we can see that steerability~$\mathcal{G}^{BD\rightarrow A}>0$~exists when ~$0.80< \eta \leq1$ ($\eta_{AB}=\eta_{SB}=\eta_{BD}=\eta_{SD}=\eta$), and steerabilities~$\mathcal{G}^{B\rightarrow A}$~and~$\mathcal{G}^{D\rightarrow A}$ are always equal to zero. If we consider transmission in a fiber with a loss of $\alpha$=0.2 dB/km ($\eta= 10^{-\alpha L/10}$), the achievable transmission distance for implementing 1sDI QSS task $L$ will be about 4.90 km. If the key is encoded on Alice's state, then it can only be unlocked by Bob and David with high accuracy if they combine measurement outcomes. A guaranteed secret key rate for providing security against eavesdropping is given by $K \geq \mathcal{G}^{BD\rightarrow A}-\text{ln}(e/2)$ \cite{YXiang2019}. As shown in Fig. \ref{figs3}, the nonzero key rate ($K > 0$) for secure QSS can be obtained within the transmission efficiency range of $0.94 < \eta \leq 1$, which means this state is a useful resource in the range of 1.34 km. This confirms that our scheme is feasible to successfully distribute various incarnations of quantum nonlocality by effectively classical means.

\subsection*{Appendix D:~~~Verification of separable conditions}

In the distribution of two-mode state, we experimentally verify the separable condition for splitting~$\hat{C}_{1}|\hat{A}\hat{B}_{0}$~by calculating the PPT value of the covariance matrix~$\sigma_{\text{A}\text{B}_{\text{0}}\text{C}_{\text{1}}}$. When the variances of the displacements added on modes~$\hat{A}_{\text{in}}$,~$\hat{C}_{\text{in}}$ are fixed to~$1.50$, $T_{1}=1/2$~and~$\eta=1$, the optimal displacement $\mathcal{F}_{B}\approx 1.24$ as given in the Table A1. The experimentally reconstructed covariance matrix is
\begin{widetext}
\begin{equation}
\sigma_{\text{A}\text{B}_{\text{0}}\text{C}_{\text{1}}}=\left( 
\begin{array}{cccccc}
2.754 & 0 & 1.296 & 0 & 0.764 & 0 \\ 
0 & 2.759 & 0 & -1.294 & 0 & -0.767 \\ 
1.296 & 0 & 3.282 & 0 & -1.296 & 0 \\ 
0 & -1.294 & 0 & 3.276 & 0 & -1.291 \\ 
0.764 & 0 & -1.296 & 0 & 2.768 & 0 \\ 
0 & -0.767 & 0 & -1.291 & 0 & 2.786%
\end{array}%
\right) 
\end{equation}
\end{widetext}
Based on the covariance matrix~$\sigma_{\text{A}\text{B}_{\text{0}}\text{C}_{\text{1}}}$, the minimum PPT values for the splittings~$\hat{A}|\hat{B}_{0}\hat{C}_{1}$,~$\hat{B}_{0}|\hat{A}\hat{C}_{1}$, and $\hat{C}_{1}|\hat{A}\hat{B}_{0}$ are~$0.701, 1.182,$~and~$1.264$, respectively. This verifies that the mode~$\hat{A}$~is inseparable with the group~$(\hat{B}_{0}\hat{C}_{1})$, but mode~$\hat{B}_{0}$ is separable from the group $(\hat{A}\hat{C}_{1})$,~mode $\hat{C}_{1}$~is separable from the group $(\hat{A}\hat{B}_{0})$, respectively. 
When modes~$\hat{C}_{1}$,~$\hat{B}_{0}$~are transmitted in lossy channels, i.e., $\eta<1$, the requirement for the separable condition will be relaxed, see Eq. \ref{eqs13}. Because the requirement for the variance of displacements corresponding to separable boundary (Eq. \ref{eqs13}) decrease with the decrease of transmission efficiency, the chosen variances of displacements ($V_{\text{dis}}=1.50$) in case of~$\eta=1$ make sure the separable conditions for smaller transmission efficiencies are also satisfied.

In order to verify the separability for splitting across~$\hat{C}_{2}|\hat{A}\hat{B}\hat{D}_{0}$~in the case of distributing three-mode entangled state, we experimentally reconstructed the
covariance matrix~$\sigma _{\text{AB}\text{C}_{\text{2}}\text{D}_{\text{0}}}$ in the case of~$V_{\text{dis}}=1.50$, $T_{1}=T_{2}=1/2,~\eta=1$ (which determine the optimal displacements $\mathcal{F}_{B}\approx 1.24$ and $\mathcal{F}_{D}\approx 1.75$),~which is given by
\begin{widetext}
\begin{equation}
\sigma _{\text{AB}\text{C}_{\text{2}}\text{D}_{\text{0}}}=\left( 
\begin{array}{cccccccc}
2.757 & 0 & 1.483 & 0 & 0.318 & 0 & 1.809 & 0 \\ 
0 & 2.753 & 0 & -1.462 & 0 & -0.312 & 0 & -1.817 \\ 
1.483 & 0 & 1.774 & 0 & 0.277 & 0 & 0.985 & 0 \\ 
0 & -1.462 & 0 & 1.777 & 0 & 0.297 & 0 & 1.061 \\ 
0.318 & 0 & 0.277 & 0 & 4.277 & 0 & 3.643 & 0 \\ 
0 & -0.312 & 0 & 0.297 & 0 & 4.251 & 0 & 3.573 \\ 
1.809 & 0 & 0.985 & 0 & 3.644 & 0 & 5.592 & 0 \\ 
0 & -1.817 & 0 & 1.061 & 0 & 3.573 & 0 & 5.606%
\end{array}%
\right) \ 
\end{equation}
\end{widetext}
Based on the covariance matrix~$\sigma _{\text{AB}\text{C}_{\text{2}}\text{D}_{\text{0}}}$, the minimum PPT values for the splittings~$\hat{A}|\hat{B}\hat{C}_{2}\hat{D}_{0}$,
~$\hat{B}|\hat{A}\hat{C}_{2}\hat{D}_{0}$,
~$\hat{C}_{2}|\hat{A}\hat{B}\hat{D}_{0}$~
and~$\hat{D}_{0}|\hat{A}\hat{B}\hat{C}_{2}$~are~$0.589$, $0.686$,~$1.177$,~and~$1.183$, respectively. From the calculating results, we can see that the second ancillary mode~$\hat{C}_{2}$ is separable from the group~$(\hat{A}\hat{B}\hat{D}_{0})$. Similarly, when modes~$\hat{C}_{2}$,~$\hat{D}_{0}$~are transmitted in lossy channels, the requirement for separable condition will also be relaxed, such that it is always satisfied at different transmission efficiencies.
\begin{figure*}[tbph]
\begin{center}
\includegraphics[width=150mm]{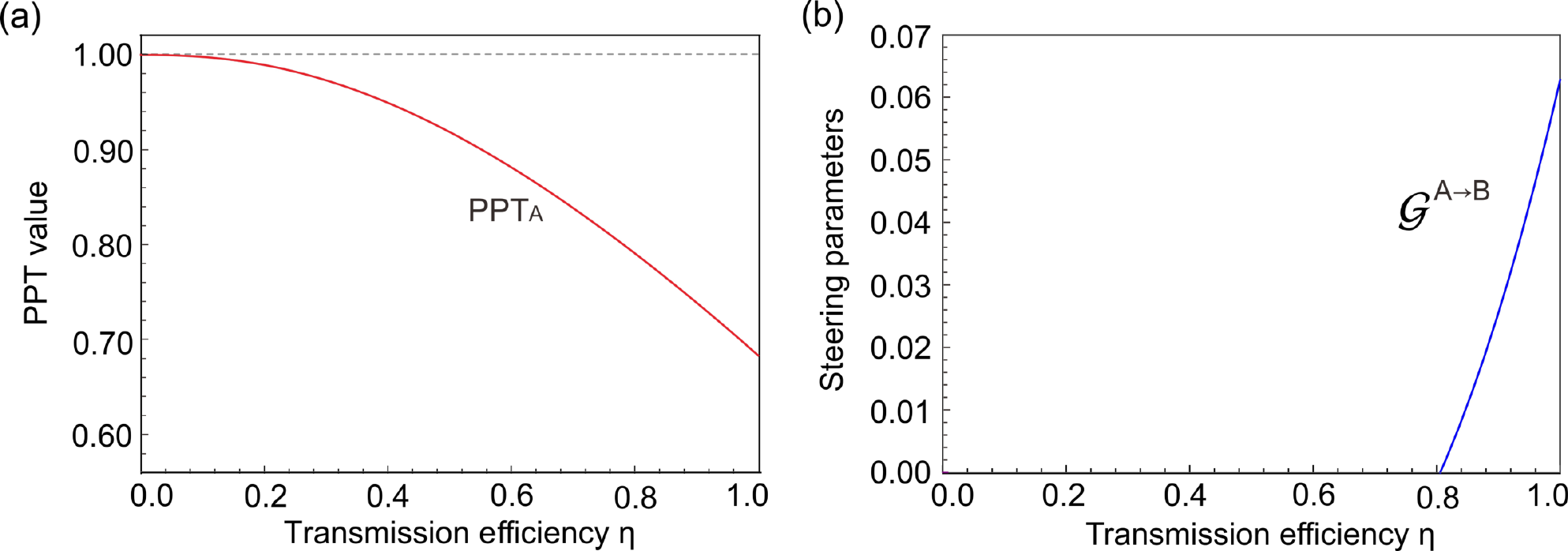}
\end{center}
\caption{Distributed entanglement and steerability when there is channel loss between the server and Alice for two users' case.} \label{figs4}
\end{figure*}
\begin{figure*}[tbph]
\begin{center}
\includegraphics[width=150mm]{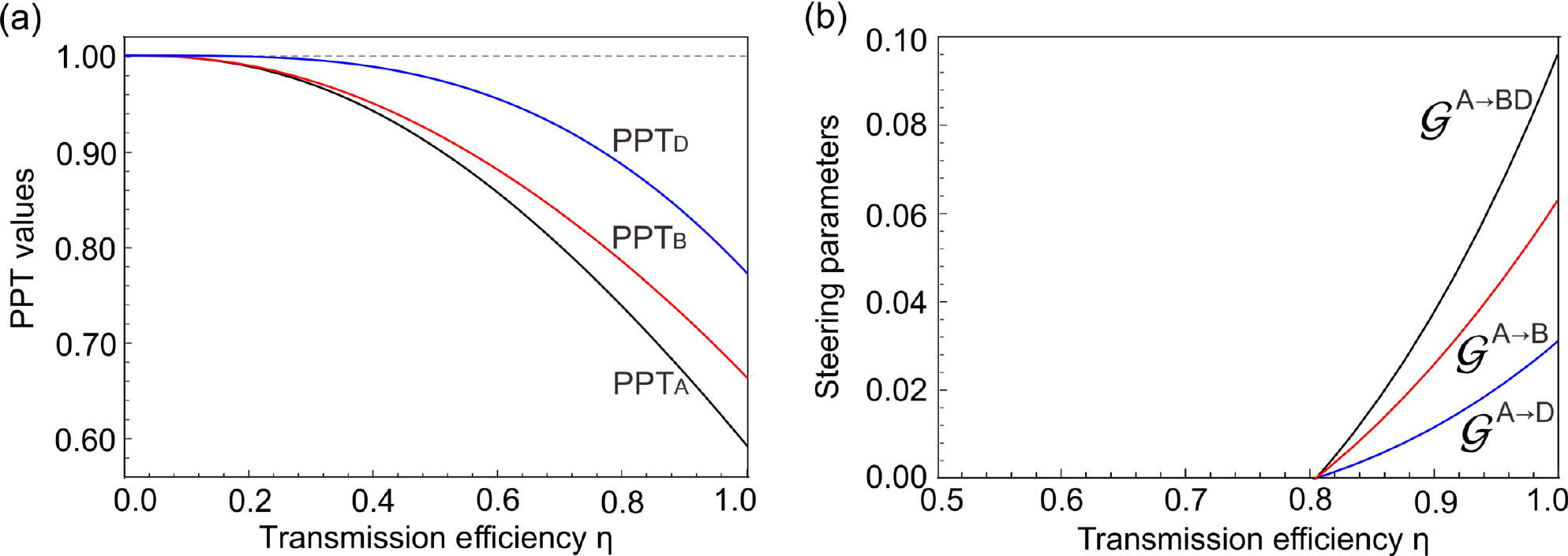}
\end{center}
\caption{Distributed entanglement and steerabilities when there is channel loss between the server and Alice for three users' case.} \label{figs5}
\end{figure*}
\subsection*{Appendix E.~~~The effect of loss between Alice and the server}

In this section, we analyze the effect of the loss between Alice and the server on the distribution of steering and entanglement, which corresponds to a network structure different from our experiment. 

In brief, the loss between Alice and the server will decrease the distance of steering distribution. In the current squeezing levels (-3/+5.5 dB), when the transmission efficiency from quantum server to Alice is assumed to be same with that from quantum server to Bob, i.e. $\eta_{SA}=\eta_{SB}=\eta_{AB}=\eta$, Gaussian entanglement is always robust against channel losses in the distribution for two users, as shown in Fig. \ref{figs4}(a). However, different from entanglement, EPR steering from Alice to Bob exists only when the transmission efficiency $\eta>0.81$. This means that the distributed steerability is more sensitive to the channel loss between Alice and server than entanglement, as shown in Fig. \ref{figs4}(b). The optimal displacement on mode $\hat{B}$ considering general losses is given by 
\begin{equation}
\mathcal{F}_{B}=\frac{2\sqrt{\eta_{SA}\eta_{AB}}[1+(V_{a}-1)\eta_{SA}]}{\sqrt{\eta_{SB}}[2+(V_{a}+V_{s}-2)\eta_{SA}]}
\end{equation}

In the case of distribution for three users, the distributed entanglement is still robust against channel losses, as shown in Fig. \ref{figs5}(a), while the steerabilities from mode $\hat{A}$ to $\hat{B}$, $\hat{D}$, the collaboration of $\hat{B}$ and $\hat{D}$ only exist when $\eta>0.81$, as shown in Fig. \ref{figs5}(b). The optimal displacements on modes $\hat{B}$ and $\hat{D}$ considering general losses are given by 
\begin{subequations}
\begin{eqnarray}
\mathcal{F}_{B}&=&\frac{2\sqrt{\eta}[1+(V_{a}-1)\eta]}{2+(V_{a}+V_{s}-2)\eta} \\
\mathcal{F}_{D}&=&\frac{2\sqrt{2}\eta[1+(V_{a}-1)\eta]}{2+(V_{a}+V_{s}-2)\eta}=\sqrt{2\eta}\mathcal{F}_{B}
\end{eqnarray}
\end{subequations}
Here, $\eta=\eta_{SA}=\eta_{SB}=\eta_{AB}=\eta_{SD}=\eta_{BD}$.

According to the above analysis, when the distances between the server and each user are the same, the distribution of entanglement is always robust against the channel losses, while the distribution of quantum steering is achievable in a limited transmission distance. For instance, the transmission distance is less than 4.9 km in a fiber channel with loss 0.2 dB/km as the transmission efficiency needs to be higher than 0.81. Note that, the robustness of steering shown in our experiment is because we considered the structure that Alice is close to the server, i.e. $\eta_{SA}=1$.
\begin{figure}[tbph]
\begin{center}
\includegraphics[width=80mm]{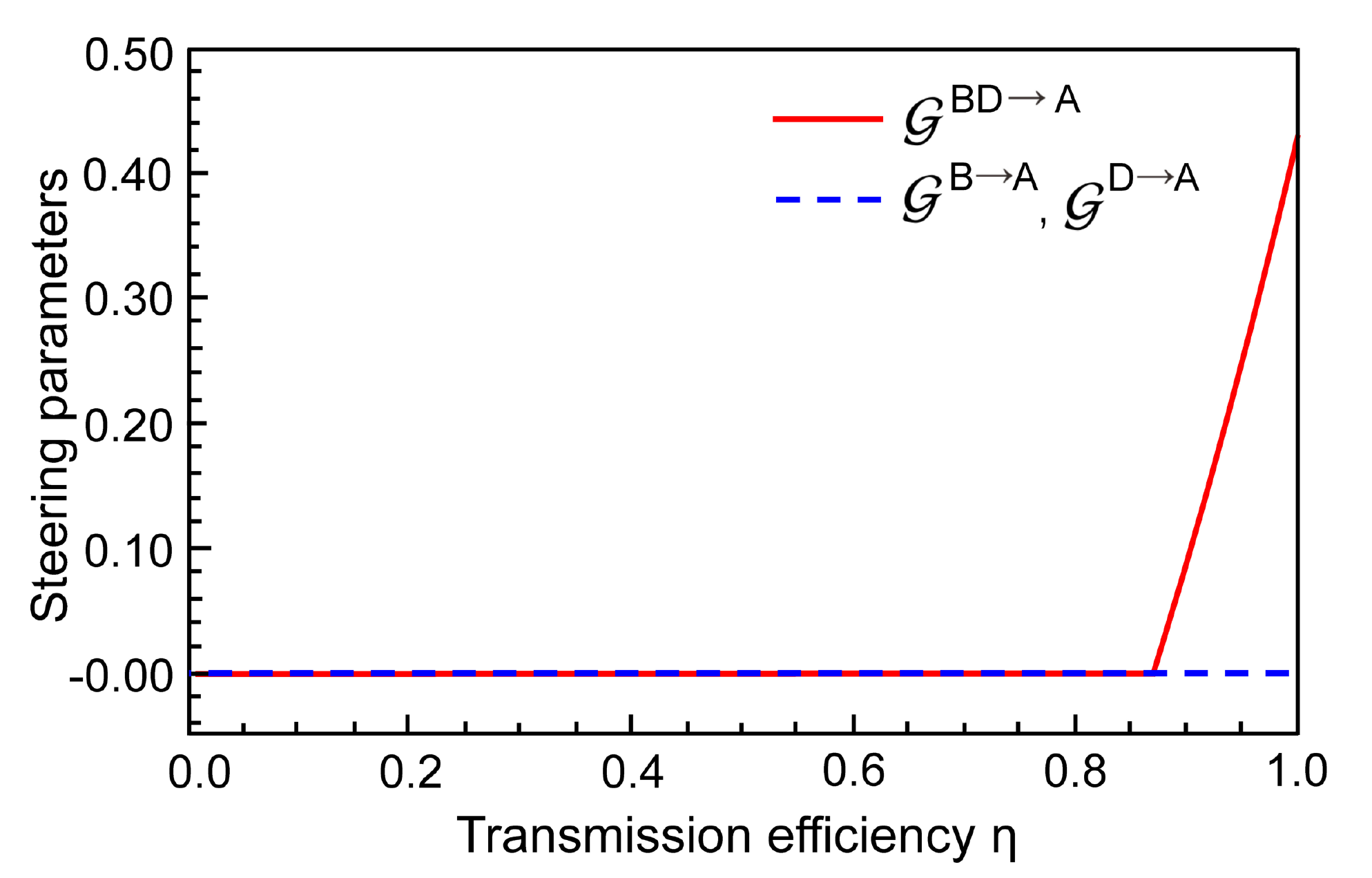}
\end{center}
\caption{The dependences of steerabilities~$\mathcal{G}^{BD\rightarrow A}$,~$\mathcal{G}^{B\rightarrow A}$~and~$\mathcal{G}^{D\rightarrow A}$~on transmission efficiency, when the channel loss between server and Alice is considered. The steerability~$\mathcal{G}^{BD\rightarrow A}>0$ when the transmission efficiency is large than $0.87$, and neither Bob nor David has ability to steer Alice.} \label{figs6}
\end{figure}

In the application of quantum secret sharing, the steerability $\mathcal{G}^{BD\rightarrow A}$ can be distributed in the range of $0.87<\eta<1$ (corresponding transmission distance is  3.02 km in a fiber channel with loss 0.2 dB/km), and $\mathcal{G}^{B\rightarrow A}$ and $\mathcal{G}^{D\rightarrow A}$ do not exist, as shown in Fig. \ref{figs6}, when the loss between Alice and the server is included.


\begin{thebibliography}{99} 

\bibitem{RevModPhys} C. Weedbrook, S. Pirandola, R. Garc\'{i}a-Patr\'{o}n, N. J. Cerf, T. C. Ralph, J. H. Shapiro, and S. Lloyd, ``Gaussian quantum information,"  Rev. Mod. Phys. \textbf{84}, 621--669 (2012).

\bibitem{rmp2020} R. Uola, A. C. S. Costa, H. C. Nguyen,  and O. G\"{u}hne, ``Quantum steering," Rev. Mod. Phys. \textbf{92},  015001 (2020).

\bibitem{review2015} D. Cavalcanti and P. Skrzypczyk,  ``Quantum steering: A review with focus on semidefinite programming," Rep. Prog. Phys. \textbf{80}, 024001 (2017).

\bibitem{reid2009} M. D. Reid, P. D. Drummond, W. P. Bowen, E. G. Cavalcanti, P. K. Lam, H. A. Bachor, U. L. Andersen, and G. Leuchs, ``Colloquium: The Einstein-Podolsky-Rosen paradox: From concepts to applications,"  Rev. Mod. Phys. \textbf{81}, 1727 (2009).

\bibitem{prxresource} R. Gallego and L. Aolita, ``Resource theory of steering," Phys. Rev. X \textbf{5}, 041008 (2015).

\bibitem{howard2007} H. M. Wiseman, S. J. Jones, and A. C. Doherty, ``Steering, entanglement, nonlocality, and the Einstein-Podolsky-Rosen paradox," Phys. Rev. Lett. \textbf{98}, 140402 (2007).

\bibitem{ExpC2009} E. G. Cavalcanti, S. J. Jones,  H. M. Wiseman, and M. D. Reid,  ``Experimental criteria for steering and the Einstein-Podolsky-Rosen paradox," Phys. Rev. A \textbf{80}, 032112 (2009).

\bibitem{EFSUV2017} A. Rutkowski, A. Buraczewski, P. Horodecki, and M. Stobi\'{n}ska, ``Quantum steering inequality with tolerance for measurement-setting errors: experimentally feasible signature of unbounded violation," Phys. Rev. Lett. \textbf{118}, 020402 (2017).

\bibitem{OneWayNatPhot} V. H\"{a}ndchen, T. Eberle, S. Steinlechner, A. Samblowski, T. Franz, R. F. Werner, and R. Schnabel, ``Observation of one-way Einstein-Podolsky-Rosen steering," Nat. Photonics \textbf{6}, 596--599 (2012).

\bibitem{ANUexp} S. Armstrong, M. Wang, R. Y. Teh, Q. Gong, Q. He, J. Janousek, H. A. Bachor, M. D. Reid, and P. K. Lam, ``Multipartite Einstein-Podolsky-Rosen steering and genuine tripartite entanglement with optical networks," Nat. Phys. \textbf{11}, 167--172 (2015).

\bibitem{prlSu} X. Deng, Y. Xiang, C. Tian, G. Adesso, Q. He, Q. Gong, X. Su, C. Xie, and K. Peng, ``Demonstration of monogamy relations for Einstein-Podolsky-Rosen steering in Gaussian cluster states," Phys. Rev. Lett. \textbf{118}, 230501 (2017).

\bibitem{noiseSu} Z. Qin, X. Deng, C. Tian, M. Wang, X. Su, C. Xie, and K. Peng, ``Manipulating the direction of Einstein-Podolsky-Rosen steering," Phys. Rev. A \textbf{95}, 052114 (2017).

\bibitem{yinprr} Y. Cai, Y. Xiang, Y. Liu, Q. He, and N. Treps, ``Versatile multipartite Einstein-Podolsky-Rosen steering via a quantum frequency comb," Phys. Rev. Research \textbf{2}, 032046(R) (2020).

\bibitem{NC2015} D. Cavalcanti, P. Skrzypczyk, G. H. Aguilar, R. V. Nery, P. H. Souto Ribeiro, and S. P. Walborn, ``Detection of entanglement in asymmetric quantum networks and multipartite quantum steering," Nat. Commun. \textbf{6}, 7941 (2015).

\bibitem{OneWayPryde} S. Wollmann, N. Walk, A. J. Bennet, H. M. Wiseman, and G. J. Pryde, ``Observation of genuine one-way Einstein-Podolsky-Rosen steering," Phys. Rev. Lett. \textbf{116}, 160403 (2016).

\bibitem{OneWayGuo} K. Sun, X.-J. Ye, J.-S. Xu, X.-Y. Xu, J.-S. Tang, Y.-C. Wu, J.-L. Chen, C.-F. Li, and G.-C. Guo, ``Experimental quantification of asymmetric Einstein-Podolsky-Rosen steering," Phys. Rev. Lett. \textbf{116}, 160404 (2016).

\bibitem{XiaoY2017} Y. Xiao, X.-J. Ye, K. Sun, J.-S. Xu, C.-F. Li, and G.-C. Guo, ``Demonstration of multisetting one-way Einstein-Podolsky-Rosen steering in two-qubit systems," Phys. Rev. Lett. \textbf{118}, 140404 (2017).

\bibitem{cvdv} A. Cavaill\`{e}s, H. Le Jeannic, J. Raskop, G. Guccione, D. Markham, E. Diamanti, M. D. Shaw, V. B. Verma, S. W. Nam, and J. Laurat, ``Demonstration of Einstein-Podolsky-Rosen steering using hybrid continuous- and discrete-variable entanglement of light," Phys. Rev. Lett. \textbf{121}, 170403 (2018).

\bibitem{steeringBEC} M. Fadel, T. Zibold, B. D\'{e}camps, and P. Treutlein, ``Spatial entanglement patterns and Einstein-Podolsky-Rosen steering in Bose-Einstein condensates," Science \textbf{360}, 409--413 (2018).

\bibitem{steeringAtomic} P. Kunkel, M. Pr\"{u}fer, H. Strobel, D. Linnemann, A. Fr\"{o}lian, T. Gasenzer, M. G\"{a}rttner, and M. K. Oberthaler, ``Spatially distributed multipartite entanglement enables EPR steering of atomic clouds," Science \textbf{360}, 413--416 (2018).

\bibitem{high2} C.-M. Li, K. Chen, Y.-N. Chen, Q. Zhang, Y.-A. Chen, and J.-W. Pan, ``Genuine high-order Einstein-Podolsky-Rosen steering," Phys. Rev. Lett. \textbf{115}, 010402 (2015).

\bibitem{high3} C.-M. Li, H.-P. Lo, L.-Y. Chen, and A. Yabushita, ``Experimental verification of multidimensional quantum steering," Opt. Commun. \textbf{410}, 956 (2018). 

\bibitem{zhangxiangdong} Q. Zeng, B. Wang, P. Li, and X. Zhang, ``Experimental high-dimensional Einstein-Podolsky-Rosen steering," Phys. Rev. Lett. \textbf{120}, 030401 (2018).

\bibitem{wangjianwei} J. Wang, S. Paesani, Y. Ding, R. Santagati, P. Skrzypczyk, A. Salavrakos, J. Tura, R. Augusiak, L. Mancinska, D. Bacco, \textit{et. al}. ``Multidimensional quantum entanglement with large-scale integrated optics," Science \textbf{360}, 285 (2018).

\bibitem{high1} Y. Guo, S. Cheng, X. Hu, B.-H. Liu, E.-M. Huang, Y.-F. Huang, C.-F. Li, G.-C. Guo, and E. G. Cavalcanti, ``Experimental measurement-device-independent quantum steering and randomness generation beyond qubits," Phys. Rev. Lett. \textbf{123}, 170402 (2019).

\bibitem{1sDIQKD_howard} C. Branciard, E. G. Cavalcanti, S. P. Walborn, V. Scarani, and H. M. Wiseman, ``One-sided device-independent quantum key distribution: security, feasibility, and the connection with steering," Phys. Rev. A \textbf{85}, 010301(R) (2012).

\bibitem{CV-QKDexp} T. Gehring, V. H\"{a}ndchen, J. Duhme, F. Furrer, T. Franz, C. Pacher, R. F. Werner, and R. Schnabel, ``Implementation of continuous-variable quantum key distribution with composable and one-sided-device-independent security against coherent attacks," Nat. Commun. \textbf{6}, 8795 (2015).

\bibitem{HowardOptica} N. Walk, S. Hosseini, J. Geng, O. Thearle, J. Y. Haw, S. Armstrong, S. M. Assad, J. Janousek, T. C. Ralph, T. Symul, H. M. Wiseman, and P. K. Lam, ``Experimental demonstration of Gaussian protocols for one-sided device-independent quantum key distribution," Optica \textbf{3}, 634 (2016).

\bibitem{YuQSS} Y. Xiang, I. Kogias, G. Adesso, and Q. He, ``Multipartite Gaussian steering: monogamy constraints and quantum cryptography applications," Phys. Rev. A \textbf{95}, 010101(R) (2017).

\bibitem{SQT15} Q. He, L. Rosales-Z\'{a}rate, G. Adesso, and M. D. Reid, ``Secure continuous variable teleportation and Einstein-Podolsky-Rosen steering," Phys. Rev. Lett. \textbf{115}, 180502 (2015).

\bibitem{SQT16_LiCM} C.-Y. Chiu, N. Lambert, T.-L. Liao, F. Nori, and C.-M. Li, ``No-cloning of quantum steering," npj Quantum Inf. \textbf{2}, 16020 (2016).

\bibitem{subchannel} M. Piani and J. Watrous, ``Necessary and sufficient quantum information characterization of Einstein-Podolsky-Rosen steering," Phys. Rev. Lett. \textbf{114}, 060404 (2015).

\bibitem{subchannel16} S.-L. Chen, C. Budroni, Y.-C. Liang, and Y.-N. Chen, ``Natural framework for device-independent quantification of quantum steerability, measurement incompatibility, and self-testing," Phys. Rev. Lett. \textbf{116}, 240401 (2016).

\bibitem{PRLSu2016} X. Su, C. Tian, X. Deng, Q. Li, C. Xie, and K. Peng, ``Quantum entanglement swapping between two multipartite entangled states," Phys. Rev. Lett. \textbf{117}, 240503 (2016).

\bibitem{otfried network} T. Kraft, S. Designolle, C. Ritz, N. Brunner, O. G\"{u}hne, and M. Huber, ``Quantum entanglement in the triangle network," arXiv: 2002.03970.

\bibitem{Miguelnetwork2020} M. Navascu\'{e}s, E. Wolfe, D. Rosset, and A. Pozas-Kerstjens, ``Genuine network multipartite entanglement," Phys. Rev. Lett. \textbf{125}, 240505 (2020).

\bibitem{qubitthe} T. S. Cubitt, F. Verstraete, W. D\"{u}r, and J. I. Cirac, ``Separable states can be used to distribute entanglement," Phys. Rev. Lett. \textbf{91}, 037902 (2003).

\bibitem{Theory2} L. Mi\v{s}ta, Jr. and N. Korolkova, ``Distribution of continuous-variable entanglement by separable Gaussian states," Phys. Rev. A \textbf{77}, 050302(R) (2008).

\bibitem{Theory3} L. Mi\v{s}ta, Jr. and N. Korolkova, ``Improving continuous-variable entanglement distribution by separable states," Phys. Rev. A \textbf{80}, 032310 (2009).

\bibitem{discord1} T. K. Chuan, J. Maillard, K. Modi, T. Paterek, M. Paternostro, and M. Piani, ``Quantum discord bounds the amount of distributed entanglement," Phys. Rev. Lett. \textbf{109}, 070501 (2012).

\bibitem{discord2} A. Streltsov, H. Kampermann, and D. Bru{\ss}, ``Quantum cost for sending entanglement," Phys. Rev. Lett. \textbf{108}, 250501 (2012).

\bibitem{qubitexp} A. Fedrizzi, M. Zuppardo, G. G. Gillett, M. A. Broome, M. P. Almeida, M. Paternostro, A. G. White, and T. Paterek, ``Experimental distribution of entanglement with separable carriers," Phys. Rev. Lett. \textbf{111}, 230504 (2013).

\bibitem{Gaussianexp1} C. E. Vollmer, D. Schulze, T. Eberle, V. H\"{a}ndchen, J. Fiur\'{a}\v{s}ek, and R. Schnabel, ``Experimental entanglement distribution by separable states," Phys. Rev. Lett. \textbf{111}, 230505 (2013).

\bibitem{Gaussianexp2} C. Peuntinger, V. Chille, L. Mi\v{s}ta, Jr., N. Korolkova, M. F\"{o}rtsch, J. Korger, C. Marquardt, and G. Leuchs, ``Distributing entanglement with separable states," Phys. Rev. Lett. \textbf{111}, 230506 (2013).

\bibitem{appnoise} M. Zuppardo, T. Krisnanda, T. Paterek, S. Bandyopadhyay, A. Banerjee, P. Deb, S. Halder, K. Modi, and M. Paternostro, ``Excessive distribution of quantum entanglement," Phys. Rev. A \textbf{93}, 012305 (2016).

\bibitem{YXiang2019} Y. Xiang, X. Su, L. Mi\v{s}ta, Jr., G. Adesso, and Q. He, ``Multipartite Einstein-Podolsky-Rosen steering sharing with separable states," Phys. Rev. A \textbf{99}, 010104(R) (2019).

\bibitem{kimble08} H. J. Kimble, ``The quantum internet," Nature (London) \textbf{453}, 1023--1030 (2008).

\bibitem{Wehner} S. Wehner, D. Elkous, and R. Hanson, ``Quantum internet: A vision for the road ahead," Science \textbf{362}, eaam9288 (2018).


\bibitem{PPT} R. Simon, ``Peres-Horodecki separability criterion for continuous variable systems," Phys. Rev. Lett. \textbf{84}, 2726 (2000).

\bibitem{Kogias2015} I. Kogias, A. R. Lee, S. Ragy, and G. Adesso, ``Quantification of Gaussian quantum steering," Phys. Rev. Lett. \textbf{114}, 060403 (2015).

\bibitem{ReidMI2013} M. D. Reid, ``Monogamy inequalities for the Einstein-Podolsky-Rosen paradox and quantum steering," Phys. Rev. A \textbf{88}, 062108 (2013).

\bibitem{NCsu} X. Su, S. Hao, X. Deng, L. Ma, M. Wang, X. Jia, C. Xie, and K. Peng, ``Gate sequence for continuous variable one-way quantum computation," Nat. Commun. \textbf{4}, 2828 (2013).

\bibitem{YZhangPRA} Y. Zhang, H. Wang, X. Li, J. Jing, C. Xie, and K. Peng, ``Experimental generation of bright two-mode quadrature squeezed light from a narrow-band nondegenerate optical parametric amplifier," Phys. Rev. A \textbf{62}, 023813 (2000).

\bibitem{OL8} X. Su, Y. Zhao, S. Hao, X. Jia, C. Xie, and K. Peng, ``Experimental preparation of eight-partite cluster state for photonic qumodes," Opt. Lett. \textbf{37}, 5178 (2012).

\bibitem{PDH} R. W. P. Drever,  J. L. Hall, F. V. Kowalski, J. Hough, G. M. Ford, A. J. Munley, and H. Ward, ``Laser phase and frequency stabilization using an optical resonator," Appl. Phys. B \textbf{31}, 97--105 (1983).

\bibitem{MCU} K. Huang, H. Le Jeannic, J. Ruaudel, O. Morin, and J. Laurat, ``Microcontroller-based locking in optics experiments," Rev. Sci. Instrum. \textbf{85}, 123112 (2014).

\end{thebibliography}
\end{document}